\documentclass[useAMS,twocolumn,usenatbib]{mn2e}

\def\OutputDriver{dvips}

\renewcommand{\vec}[1]{ {\bmath #1} } 

\usepackage[\OutputDriver]{graphicx}
\usepackage{amssymb}
\usepackage{amsmath}
\usepackage{bm}

\begin{document}

\title[Intracluster stars in simulations with
AGN feedback]
{Intracluster stars in simulations with
AGN feedback}
\author[Ewald Puchwein, Volker Springel, Debora Sijacki, \& Klaus
Dolag]{
Ewald Puchwein$^1$,
Volker Springel$^1$,
Debora Sijacki$^2$,
and
Klaus Dolag$^1$
\\$^1$Max-Planck-Institut
f{\"u}r Astrophysik, Karl-Schwarzschild-Stra{\ss}e 1, 85741 Garching,
Germany
\\$^2$Kavli Institute for Cosmology, Cambridge and Institute of
Astronomy,
Madingley
Road, Cambridge, CB3 0HA, United Kingdom
}
\date{\today}
\maketitle

\begin{abstract}
We use a set of high-resolution hydrodynamical simulations of
clusters of galaxies to study the build-up of the intracluster light
(ICL), an interesting and likely significant component of their total
stellar mass. Our sample of groups and clusters includes AGN feedback
and is of high enough resolution to accurately resolve galaxy
populations down to the smallest galaxies that are expected to
significantly contribute to the
stellar mass budget. We describe and test four different methods to
identify the ICL in cluster simulations, thereby allowing us to assess
the reliability of the measurements. For all of the methods, we
consistently find a very significant ICL stellar fraction ($\sim45\%$)
which exceeds the values typically inferred from observations. However,
we show that this result is robust with respect to numerical resolution
and integration accuracy, remarkably insensitive to changes in the star
formation model, and almost independent of halo mass. It is also almost
invariant when black hole growth is included, even though AGN feedback
successfully prevents excessive overcooling in clusters and leads to a
drastically improved agreement of the simulated cluster galaxy
population with observations. In particular, the luminosities of
central cluster galaxies and the ages of their stellar populations are
much more realistic when including AGN. In  the light of these findings,
it appears challenging to construct a simulation model that
simultaneously matches the cluster galaxy population and at the same
time produces a low ICL component. We find that intracluster stars are
preferentially stripped in a cluster's densest region from massive
galaxies that fall into the forming cluster at $z>1$. Surprisingly, some
of the intracluster stars also form in the intracluster medium inside
cold gas clouds that are stripped out of infalling galaxies.
\end{abstract}


\section{Introduction}
\label{sec:Introduction}

In optical observations, galaxy clusters appear as concentrations
of galaxies on the sky. However, deep exposures reveal that not all
the emission comes from stars that reside in the cluster's member
galaxies.  Instead, there is also a smoothly distributed stellar
component which is typically peaked around the cluster's central
galaxy, but extends to large radii. Due to its low surface brightness,
observations of this intracluster light (ICL) component are difficult,
resulting in a significant uncertainty in the current observational
constraints of the amount of ICL present in clusters.

\cite{Lin2004} give an overview of the fractions of intracluster
stars reported in the literature for different clusters and groups
(see their Table 2). The reported values span a huge range from
$\sim2\%$ to $\sim50\%$.  Instead of analyzing very deep exposures of
individual objects, \cite{Zibetti2005} stacked hundreds of cluster
images from the Sloan Digital Sky Survey and masked all satellite
galaxies, allowing them to obtain a very deep exposure of the
brightest cluster galaxy (BCG) and the ICL of an average galaxy
cluster.  For a mean cluster mass of $7-8\times10^{13} M_\odot$ they
find that about $\sim22\%$ of the stars reside in the BCG, while
roughly $\sim11\%$ are intracluster stars within a fixed aperture of
500 kpc.  \citet{Gonzalez2005,Gonzalez2007} describe another study for
which a significant number of clusters and groups were observed.  By
analyzing the observed systems individually, they found that both the
total stellar fractions as well as the relative fractions of stars
residing in the BCG+ICL component strongly increase with decreasing
halo mass. It was also concluded that intracluster stars are important
for the baryon budget of clusters and groups.

The latter is particularly interesting in the light of the recent
findings that the baryon fractions measured in clusters seem to be
significantly lower than those inferred from the most current
cosmological constraints \citep[see e.g.][]{McCarthy2007}.  However,
no mechanism that can efficiently segregate baryons from dark matter
on the scale of massive galaxy clusters is known. On the other hand,
if there is indeed a significant component of intracluster stars,
correctly accounting for them may relax the reported tensions.

The distribution and abundance of intracluster stars were also
investigated based on cosmological hydrodynamical simulations of
galaxy cluster formation that included radiative cooling and heating, as
well as models for star formation and supernova feedback
\citep[][]{Murante2004,Willman2004,Sommer-Larsen2005,Murante2007,
Dolag2009}. The origin of intracluster stars was studied
in particular detail in \citet{Murante2007} finding an ICL
fraction that rises with halo mass from $\sim10\%$ at the group scale
to $\sim30\%$ for massive clusters. In their simulations, about half
of the intracluster stars came from galaxies associated with the
merger tree of the BCG, and most intracluster stars were liberated
from their former host galaxies during merger events after redshift
$z=1$. Using similar simulations but a different method to identify
the ICL, \cite{Dolag2009} obtained average ICL fractions of around
$\sim33\%$, with a significant scatter on the group scale. No trend of
the mean ICL fraction with halo mass was detected when their new ICL
definition was used.

A major problem of most hydrodynamical cluster simulations thus far
has been that they suffered from excessive ``overcooling'' within the
densest cluster regions, where the gas cooling times are short. As a
result, an unrealistically large fraction of cold gas and consequently
a large amount of stars in clusters formed from the strong cooling
flows.  This in turn leads to central galaxies that are too bright and
too blue, as well as extremely high total stellar fractions. It is
widely believed that some source of non-gravitational energy input,
like heating by active galactic nuclei (AGN), is necessary to offset
the cooling in cluster cores and to make simulated clusters compatible
with observations. In \cite{Sijacki2007} and \cite{Puchwein2008}
it was shown that including a model for AGN feedback in such
simulations, indeed, strongly improves the agreement between simulated
and observed X-ray properties of clusters and groups \citep[see
also][]{McCarthy2009}.

In this work, we explore whether AGN feedback can also resolve
discrepancies between simulated and observed cluster galaxy
populations, and in particular, between the amount of intracluster
light. Our high-resolution simulation set is well suited to
investigate the origin of the ICL because for the first time a large
simulated sample of galaxy clusters and group is available which can
resolve the cluster galaxy populations down to very low galaxy masses
and at the same time includes a successful model for the growth
and feedback activity of AGN.

This paper is structured as follows. In Section~2 we describe our
simulation set, the numerical models used, and the analysis methods we
employ for measuring the ICL component in the simulations. We also
summarize the tests we have carried out to investigate the robustness
of our results with respect to numerical parameters and variations in
the star formation model. In Section~3, we present our main results
on the origin and amount of ICL found in our simulations, as well as on
the properties of the simulated cluster galaxy populations.
Finally, we summarize our findings and give our conclusions
in Section~4.

\section{Methods}
\label{sec:methods}

We have carried out high-resolution cosmological hydrodynamical
simulations of a large sample of galaxy clusters and groups, including
a treatment of star formation and feedback processes.  We will here
mainly focus on analyzing the distribution of stars in the formed
objects. To this end we develop and intercompare several distinct
methods to measure the following components in our simulated clusters:
the central galaxy (BCG),\footnote{The central galaxy, i.e.~the galaxy
  closest to the minimum of the cluster potential well, is typically
  also the brightest cluster galaxy, except for very rare
  exceptions. For simplicity, we always refer to the central galaxy of
  a simulated cluster or group when we use the abbreviation BCG.}  the
satellite galaxies (i.e. all other cluster member galaxies), and the
ICL. We then investigate the properties of these components and their
relation to each other.

\subsection{The simulations}
\label{sec:simulations}

Our simulated galaxy cluster and group sample is described in detail
in \cite{Puchwein2008}. In this work we use the 16 most massive
objects of the sample. They approximately uniformly cover the mass
range from $M^\text{crit}_{200} = 2\!\times\!10^{13} M_\odot$ to $1.5
\! \times \! 10^{15} M_\odot$.\footnote{Throughout this paper,
  spherical overdensity masses and radii will be denoted by the
  symbols $M$ and $r$, where we use the superscripts "crit" or "mean"
  to indicate whether the overdensity is with respect to the critical
  density of the Universe or the mean matter density of the Universe
  at the cluster redshift, respectively. The subscripts 200 and 500
  indicate how many times larger the mean density inside the
  corresponding spherical region is compared to the chosen reference
  density.}

In brief, we have selected dark matter halos from the Millennium
simulation \citep{Springel2005a} at $z=0$ and resimulated them at
higher mass and force resolution, including a gaseous component and
accounting for hydrodynamics, radiative cooling, heating by a UV
background, star formation and supernovae feedback. For each halo, at
least two kinds of resimulations were performed. One containing the
physics just described, and an additional one that also included a
model for feedback from active galactic nuclei (AGN). We note that the
original halo selection was only based on mass and otherwise
random. New initial conditions for the resimulations were created by
populating the Lagrangian region of each halo in the original initial
conditions with more particles and adding additional small-scale
power, as appropriate. At the same time, the resolution has been
progressively reduced in regions that are sufficiently distant from
the forming halo. Gas has been introduced into the high-resolution
region by splitting each parent particle into a gas and a dark matter
particle.

We have adopted the same flat $\Lambda$CDM cosmology as in the parent
Millennium simulation, namely $\Omega_{\rm m}=0.25$, $\Omega_{\rm
  \Lambda}=0.75$, $h=0.73$, $n_s=1$ and $\sigma_8=0.9$. A baryon
density of $\Omega_{\rm b}=0.04136$ has been chosen for consistency
with the cosmic baryon fraction inferred from current cosmological
constraints \citep{Komatsu2008}.

The simulations were run with the {\small GADGET-3} code \citep[based
on][]{Springel2005c}, which employs an entropy-conserving formulation of
smoothed particle hydrodynamics (SPH). Radiative cooling and heating was
calculated for an optically thin plasma of hydrogen and
helium, and for a time-varying but spatially uniform UV background.
Star formation and supernovae feedback were modelled with a subresolution
multi-phase model for the interstellar medium as in \cite{Springel2003}.
Black hole growth and associated feedback processes were followed as in
\cite{Springel2005b} and \cite{Sijacki2007}.

Table~\ref{tab:resolutions} summarizes the different mass and force
resolutions that were used in our resimulations. For all halos with
$M^\text{crit}_{200} \le 1.5\times10^{14} \, h^{-1} M_\odot$ simulations
with a
zoom factor of 3 (see Table \ref{tab:resolutions}) have been
performed, while all more massive cluster were resimulated only with a
zoom factor of 2. For some objects several resimulations have been
performed, covering a wide range in mass and force resolution. This
allows us to assess the convergence of the simulations.

\begin{table}
\begin{center}
\begin{tabular}{ccccc}
\hline
zoom&softening&\multicolumn{3}{c}{particle mass $[h^{-1} M_\odot]$} \\
factor & {\small $[h^{-1}\textrm{kpc}]$} & DM & gas & stars\\
\hline
1 & 7.5  & $8.3\times10^8$ & $1.8\times10^8$ & $8.9\times10^7$ \\
2 & 3.75 & $1.1\times10^8$ & $2.1\times10^7$ & $1.0\times10^7$ \\
3 & 2.5  & $3.1\times10^7$ & $6.2\times10^6$  & $3.1\times10^6$ \\
4 & 1.875 & $1.3\times10^7$ & $2.8\times10^6$ & $1.4\times10^6$ \\
\hline
\end{tabular}
\end{center}
\caption{Force and mass resolutions used in the cluster and
group simulations. The physical (Plummer-equivalent)
softening lengths that were used for redshifts $z<5$ are listed in
Column 2. For $z>5$,
comoving softenings of $45\,h^{-1}{\rm kpc}$, $22.5\,h^{-1}{\rm kpc}$,
$15\,h^{-1}{\rm kpc}$, and $11.25\,h^{-1}{\rm kpc}$ were used for zoom
factors 1, 2, 3, and 4, respectively. The masses of the dark matter
(DM), gas and star particles in the simulations are tabulated in Columns
3-5.}
\label{tab:resolutions}
\end{table}

\subsection{Identifying intracluster stars and cluster galaxies}
\label{sec:identify_icl}

\begin{figure}
\centerline{\includegraphics[width=\linewidth]{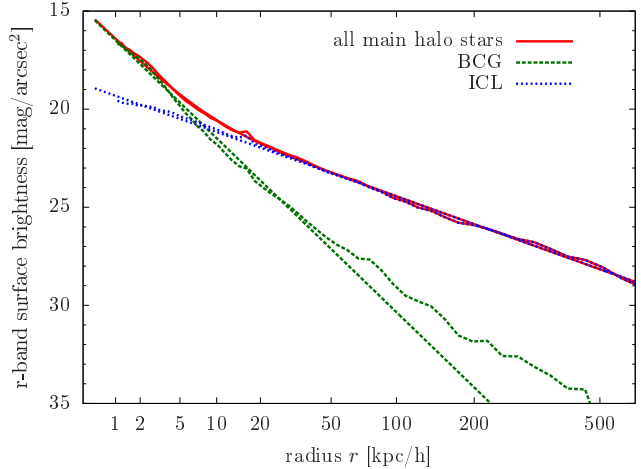}}
\caption{r-band surface brightness profiles of main halo stars of a
$2.2\times 10^{14} \, h^{-1} M_\odot$ cluster. The {\it red solid}
curves show
the surface brightness profile of all main halo stars and its 
two-component de Vaucouleurs fit (smooth curve). The straight {\it green
dashed} and {\it blue dotted} lines indicate the two individual de
Vaucouleurs components corresponding to the BCG and ICL. The additional
{\it green dashed} and {\it blue dotted} curves indicate the surface
brightness profiles of the BCG and ICL as found by {\it Method 4}.
For this cluster the results of {\it Method 2} and {\it Method 4} agree
very well. The $r$-axis has been chosen to be linear in $r^{1/4}$. The
simulation has been performed without AGN feedback.}
\label{fig:sb_profiles}
\end{figure}

\begin{figure}
\centerline{\includegraphics[width=\linewidth]{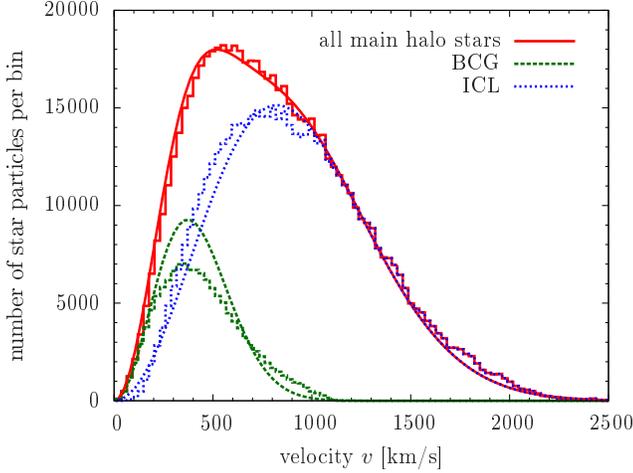}}
\caption{Velocity distribution of main halo stars of a
$2.2\times 10^{14} \, h^{-1} M_\odot$ cluster. The {\it red solid}
histogram
shows the velocity distribution of all main halo stars. The smooth {\it
red solid} curve is the two component Maxwell-Boltzmann fit used in {\it
Method 3} and {\it 4}. The smooth {\it green dashed} and {\it blue
dotted} curves indicate the individual Maxwell-Boltzmann distributions
corresponding to the BCG and ICL. The {\it green dashed} and {\it
blue dotted} histograms show the velocity distributions of the BCG and
ICL as identified by {\it Method 4}. The
simulation has been performed without AGN feedback.}
\label{fig:vel_dist}
\end{figure}

Starting from the particle distribution in the simulated clusters, we
aim to identify the cluster's central galaxy, all satellite member
galaxies and the ICL component. As a first step, we run a
friends-of-friends (FoF) group finder with a linking length of 0.2 in
units of the mean interparticle distance, as well as the {\small
  SUBFIND} substructure finder \citep{Springel2001}. The FoF algorithm
is applied only to the high-resolution dark matter particles, with the
gas and stars linked to their nearest dark matter particle. This
avoids biases in the group catalogue due to the very different spatial
distribution of dark matter and baryonic particles.  We then use a
version of {\small SUBFIND} that was specifically adapted to work
robustly with hydrodynamical simulations \citep{Dolag2008} for
decomposing each FoF-group into a main halo and self-bound
substructures. We consider the stellar content of each self-bound
substructure of the cluster to be a satellite galaxy. All star
particles that are part of the main halo are assumed to belong either
to the BCG or to the ICL component. It is, however, not
straightforward to unambiguously decide to which of these two
components such a star particle should be assigned.  In the following,
we therefore employ several independent methods for making a
distinction between the BCG and the ICL in simulated clusters. These
are:

\begin{itemize}

\item {\it Method 1: A cut-off radius around the central galaxy.} The
  simplest possible approach is to just cut off the BCG at a specific
  radius $r_\textrm{cut}$. However, the size of BCGs correlates with
  cluster mass. We therefore choose to scale $r_\textrm{cut}$ with
  cluster mass. We obtain an appropriate scaling relation by combining
  an empirical relation between cluster mass and BCG luminosity
  \citep{Popesso2007} with a relation between BCG luminosity and
  half-light radius $r_\textrm{e}$ \citep{Bernardi2007}. We decide to
  cut at $r_\textrm{cut} = 3 \times r_\textrm{e}$, yielding
\begin{equation}
r_\textrm{cut} = 27.3 \,\, h^{-1}\textrm{kpc} \times
\left(\frac{M^\text{crit}_{200}}{10^{15} \, h^{-1}
M_\odot}\right)^{0.29}.
\end{equation}
When using this method, all star particles of the main halo that are
within the cut-off radius are considered to be part of the BCG, while
all main halo star particles outside this radius are considered to be
intracluster stars.

\item {\it Method 2: Analysis of the surface brightness profiles.} A
  de Vaucouleurs profile \citep{deVaucouleurs1948} is usually a good
  fit to the surface brightness profile of an elliptical galaxy. For
  cD galaxies, however, one often finds a light excess at large radii
  with respect to such a fit to the inner region
  \citep{Schombert1986}.  This light excess can be attributed to
  intracluster stars.

The complete surface brightness profile of BCG and ICL can typically
be well fit by the sum of two de Vaucouleurs profiles \citep[see
  e.g.][]{Gonzalez2005}. We perform such a two-component fit to the
surface brightness profile of all stars residing in a simulated
cluster's main halo, as found by {\small SUBFIND}, and consider the
component with the smaller effective radius as the BCG and the
component with the larger effective radius as the ICL. This procedure
is illustrated for a cluster with $M^\text{crit}_{200} = 2.2\times
10^{14} \, h^{-1} M_\odot$ in Fig.~\ref{fig:sb_profiles}. We note that
we
obtain projected mass profiles for the BCG and the intracluster stars
by assuming that in each radial bin both components have the same
mass-to-light ratio. In order to compute surface brightness
profiles we use the stellar population synthesis model library GALAXEV
\citep{Bruzual2003} to assign a r-band luminosity to each star
particle according to its mass, age and metallicity.

\item {\it Method 3: Analysis of the stellar velocity distribution.}
  In relaxed simulated clusters, the distribution of the velocities
  $v$ of the main halo stars can be well fit by a sum of two
  Maxwell-Boltzmann distributions \citep[see][]{Dolag2009}. Here $v$
  is given by $v=|\vec{v}_\textrm{star}-\vec{v}_\textrm{clus}|$, where
  $\vec{v}_\textrm{star}$ is the velocity of the star particle and
  $\vec{v}_\textrm{clus}$ is the velocity of the centre of mass of the
  cluster's main halo. In this method, we perform such a fit for each
  cluster and define the BCG mass as the mass of the stars in the
  Maxwell-Boltzmann component with the lower velocity dispersion,
  while we associate the higher dispersion component with the
  ICL. This is illustrated in Fig.~\ref{fig:vel_dist} for a cluster
  with mass $M^\text{crit}_{200} = 2.2\times 10^{14} \, h^{-1} M_\odot$.

\item {\it Method 4: Analysis of the stellar velocity distribution and
  binding energies.} While {\it Method 3} allows an estimate of the
  total stellar BCG mass, it does not assign individual star particles
  to either the BCG or ICL component, as it yields, for each velocity
  bin, only the fraction of stars in each component. However, by also
  including the binding energies of the star particles in the
  analysis, the method can be extended to allow such an individual
  assignment, as discussed in full detail in \citet{Dolag2009}.  

In short, we use this approach in the following way. We start with the
two component Maxwell-Boltzmann fit obtained from {\it Method 3}. Then
we use an iterative algorithm that assigns star particles according to
their binding energy either to the BCG or to the ICL, with the
iteration repeated until the BCG and ICL velocity distributions
roughly match the fitted Maxwell-Boltzmann components.  We start the
iteration with some value $r_\textrm{grav}$ for the size of the BCG's
gravitating mass distribution. Then, the gravitational potential of
all simulation particles with a distance $r < r_\textrm{grav}$ from
the cluster centre is computed. All main halo star particles that are
bound in this potential are assigned to the BCG in this iteration
step, while unbound main halo stars are assigned to the ICL. Next, the
velocity dispersion of the ICL component is calculated and compared to
the velocity dispersion of the two Maxwell-Boltzmann fits. If the
velocity dispersion of the ICL is too large, $r_\textrm{grav}$ is
reduced to unbind more of the slowly moving particles.  Otherwise,
$r_\textrm{grav}$ is increased. We stop the iteration once the ICL
velocity dispersion agrees with the velocity dispersion of the higher
dispersion Maxwell-Boltzmann component to within 1\%.
Fig.~\ref{fig:vel_dist} displays the resulting velocity distributions
of the BCG and ICL components obtained in this way and compares them
to the corresponding Maxwell-Boltzmann fits.
\end{itemize}

In Fig.~\ref{fig:sb_profiles}, we compare the dynamical definition of
BCG and ICL obtained with {\it Method 4} to the analysis of the
surface brightness profiles used in {\it Method 2}. For the shown
cluster the agreement between these two completely different methods
is remarkably good. However, in some other clusters the differences
are admittedly larger (see also Fig.~\ref{fig:component_fractions}).

We note that we can obtain a BCG and ICL mass by all of these four
methods. However, only {\it Method 1} and {\it Method 4} allow us to
individually assign each main halo star particle to one of these
two components.

\subsection{Robustness with respect to numerical resolution
and integration accuracy}
\label{sec:res_studies}

Before we apply the methods introduced above to our full cluster and
group sample, we investigate whether our satellite galaxy, BCG and
ICL stellar masses are robust with respect to numerical resolution by
analysing three halos that were simulated at several different mass and
force resolutions. We also test our integration accuracy with one
further cluster simulation where we used two times smaller
time steps and a higher force accuracy setting in our tree
code.\footnote{In this run, the opening angle for nodes in the tree
  part of the force computation is a factor $\sim 1.6$ smaller.}

\begin{figure}
\centerline{\includegraphics[width=\linewidth]{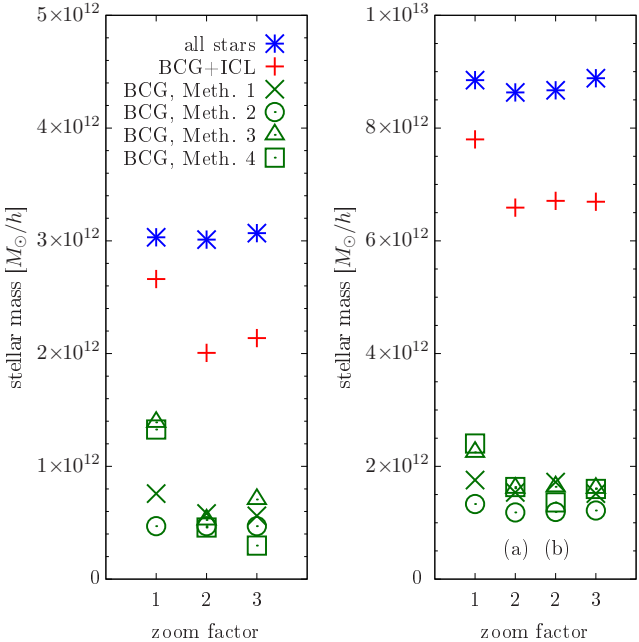}}
\caption{BCG, BCG+ICL and total stellar mass found within
$r^\text{crit}_{500}$ in a $4.4 \times 10^{13} \, h^{-1} M_\odot$ group
({\it
left panel}) and a $1.5 \times
10^{14} \, h^{-1} M_\odot$ cluster ({\it right panel}) simulated without
AGN at different numerical
resolutions. Results are shown for zoom factor 1, 2, and 3
resolutions (see Table \ref{tab:resolutions}) as well as for the four
different methods for determining a BCG mass (see
Sect.~\ref{sec:identify_icl}). In the  {\it right panel}, two zoom
factor 2 simulations are shown, one with the standard parameters ({\it
a}) and one with smaller time steps and a more accurate force
computation ({\it b}). Results are well converged for zoom factors 2
and larger.}
\label{fig:res_comp}
\end{figure}

Fig.~\ref{fig:res_comp} shows the masses of all stars, of all main
halo stars (i.e. BCG+ICL), and of the BCG in these test simulations
for a group with $M^\text{crit}_{200} = 4.4 \times 10^{13} \,
h^{-1} M_\odot$
and a cluster with mass $M^\text{crit}_{200} = 1.5 \times 10^{14} \,
h^{-1} M_\odot$. We computed these masses within $r^\text{crit}_{500}$,
based on resimulations with zoom factors 1, 2, and 3, without
inclusion of AGN physics. The BCG masses obtained with all the four
different methods introduced in Sect.~\ref{sec:identify_icl} are
shown. The total stellar masses are well converged even in our lowest
resolution simulations without AGN.  However, the mass of stars in
satellite galaxies, which is given by the difference between the
results shown for all stars and for the BCG+ICL component, is not
fully converged at the resolution of zoom factor 1. On the other hand,
the results for zoom factors 2 and 3 for the BCG+ICL and the satellite
galaxy mass agree very well, indicating that for zoom factor 2 and
larger good convergence is achieved.

The BCG masses that are inferred with the different analysis methods
can sometimes slightly differ from each other. This will be discussed
in more detail in Section~\ref{sec:bcg_icl_sat_frac}. Here, we want to
assess the numerical convergence of the results obtained by each
individual method. The BCG masses obtained by using a cutoff radius
({\it Method 1}) or the surface brightness profile analysis ({\it
  Method 2}) are already roughly converged even at the zoom factor 1
resolution. The results for {\it Method 3} and {\it Method 4} on the
other hand, which both rely on fitting the velocity distribution of
main halo stars, seem to be somewhat more easily affected by low
numerical resolution, because the masses obtained for zoom factor 1
are larger than those obtained from the higher resolution
runs. However, at least for the group shown in the left panel of
Fig.~\ref{fig:res_comp}, this seems partly due to a slightly different
timing of a rather massive subhalo that is just passing by close to
the cluster center in the low resolution simulation, so it is not
really clear how generic this effect is. In any case, we find that the
BCG masses have robustly converged at zoom factors 2 and larger.

However, we also find that the resolution requirements to obtain
converged results for the total stellar mass in runs with AGN feedback
are somewhat higher. This is because at higher resolution star
formation shifts to smaller halos, which contain smaller black holes
that are less effective in suppressing star formation by their
feedback. As a consequence, the total stellar mass in our simulated
clusters is not nearly as well converged at the resolution of
zoom factor 1 as in our runs without AGN.  It is, however, reasonably
converged at zoom factor 2 (stellar mass about $\sim 20\%$ low) and well
converged at zoom factor 3 resolution (about $\sim 6\%$ low).

The results of the run with smaller time steps and higher force
accuracy agree extremely well with those obtained form the
corresponding run with our standard numerical parameters (see right
panel of Fig.~\ref{fig:res_comp}).  We hence conclude that the orbit
integration is sufficiently accurate in our simulations.  This is
important and reassuring, as an insufficient integration accuracy may
lead to an artificially enhanced stripping rate of stars out of
cluster galaxies.

\begin{figure}
\centerline{\includegraphics[width=\linewidth]{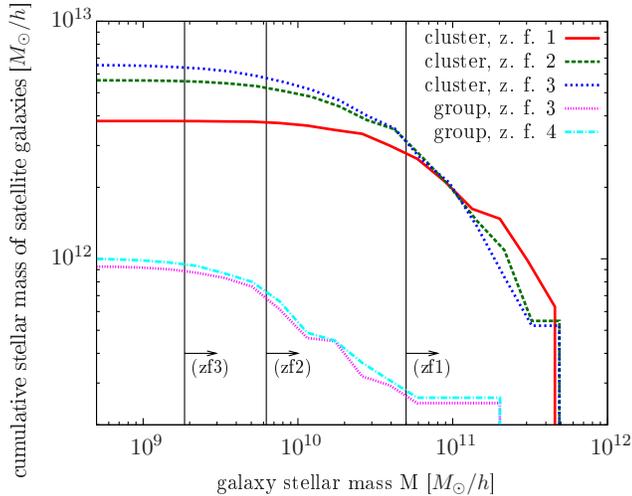}}
\caption{Combined stellar mass of all satellite galaxies with
individual stellar masses exceeding $M$ as a function of galaxy stellar
mass $M$. Results are shown for zoom factor 1, 2 and 3 simulations of a
$M^\text{crit}_{200} = 1.5 \times 10^{14} \, h^{-1} M_\odot$ cluster and
zoom factor 3 and 4
simulations of a $M^\text{crit}_{200} = 3\times 10^{13} \, h^{-1}
M_\odot$ group. No AGN have
been 
included in these runs. The vertical lines, arrows and, labels {\it (zf1)}, {\it (zf2)}, and {\it (zf3)} indicate
the galaxy mass ranges in which the stellar mass in satellite galaxies
is approximately converged in zoom factor 1, 2, and 3 simulations,
respectively.}
\label{fig:cum_massfuns}
\end{figure}

In Fig.~\ref{fig:cum_massfuns}, we explore the convergence of the
properties of the satellite galaxy population in more detail. Here the
combined stellar mass of all satellite galaxies whose individual
stellar masses exceed $M$ is shown as a function of $M$ for
simulations at different resolution. Comparing the curves obtained
from simulations with zoom factors 1, 2, 3, and 4, we conclude that
the satellite galaxy population is reasonably converged for stellar
masses larger than $\sim 5 \times 10^{10} \, h^{-1} M_\odot$, $\sim 6
\times
10^9 \, h^{-1} M_\odot$ and $\sim 2 \times 10^9 \, h^{-1} M_\odot$, at
zoom factors 1,
2 and 3, respectively. In other words, the formation, merger and tidal
disruption rates of galaxies above these masses should be robust with
respect to numerical resolution. Using the double Schechter fit to the
observed galaxy stellar mass function from \cite{Baldry2008}, we
estimate that $\sim69\%$, $\sim17\%$ and $\sim9\%$ of the total
stellar mass are contained in galaxies with individual stellar masses
below these resolution limits. Thus, at zoom factor 1 the majority of
stars resides in poorly resolved galaxies. This explains why we obtain
lower values for the mass in satellite galaxies in these runs (see
Fig.~\ref{fig:res_comp}). On the other hand, in the runs with zoom
factors 2 and 3, the vast majority of stars resides in galaxies whose
abundance is numerically converged. We thus do not expect a
significant spurious contribution to the ICL component form
underresolved galaxies in these simulations.

Overall, these test simulations indicate that the distribution of
stars in our cluster resimulations with zoom factors 2 and 3 are robust
with respect to mass and force resolution, except for maybe a slight
underestimation of the total star formation efficiency in zoom factor
2 runs that include AGN. Importantly, we have also verified that our
results are invariant when the time step size is reduced or the 
accuracy of the gravitational force computation is improved.

\subsection{Dependence on the star formation model}
\label{sec:sf_model_changes}

We also investigated how sensitive our results are to changes in the
parameters of the star formation model. More precisely, we performed
two test simulations, one with a higher gas density threshold for star
formation, and another one in which we assumed a larger timescale for
the conversion of gas into stars.

The idea behind using a higher gas density threshold was that this
might lead to smaller and denser galaxies, which should then be less
prone to tidal stripping. This in turn could reduce the amount of ICL
in the simulated clusters. In one of our tests, we therefore increased
the threshold density for star formation by a factor of $\sim 4$
compared to the original simulation.  Within the subresolution model
of \citet{Springel2003} for the regulation of star formation this was
achieved by increasing the `supernova temperature' $T_\text{SN}$ from
$10^8 \text{K}$ to $4\times 10^8 \text{K}$, as well as the supernova
evaporation factor $A_0$ from 1000 to 4000, while keeping the gas
consumption timescale the same.  However, the analysis of this
simulation yielded very similar results to the original run. The
fraction of stars in the BCG+ICL component changed by only about
$1\%$.

Varying instead the timescale for star formation and thereby shifting
the peak of the star formation activity to a different stage in the
assembly history of the cluster might also affect the ICL component.
In particular, since mergers play an important role in liberating
stars from their host galaxies \citep[see][]{Murante2007}, the
efficiency of this mechanism for creating ICL stars may depend on the
star formation history of the cluster.  The idea behind our second
test of the star formation model was therefore to increase the
timescale $t_0^*$ for the conversion of gas into stars from 2.1 Gyrs
to 8.4 Gyrs. Keeping everything else equal, this required again to
set $T_\text{SN} = 4\times 10^8 \text{K}$ and $A_0 = 4000$ to leave
the gas density threshold at its original value, while $t_0^*$ itself
was increased by a factor of 4.  Using this larger gas consumption
timescale indeed shifts the peak of the star formation history to much
later times and broadens it at the same time. It also results in more
star formation at low redshift. However, also for this simulation we
find remarkably little difference in the galaxy population and the ICL
compared to the original run. Despite the substantial change in the
star formation history, the fraction of stars in the BCG+ICL component
increased by only $4\%$.

Overall, these tests show that the fraction of stars that get unbound
from their host galaxies and become intracluster stars is remarkably
insensitive to the detailed parameters of the star formation
model.

\section{Results}
\label{sec:results}

We now turn to the presentation of our primary results, obtained by
applying the methods introduced in Sect.~\ref{sec:identify_icl} to our
full sample of simulated clusters.\footnote{Unless indicated
  otherwise, we use the redshift $z=0$ outputs.} In particular, we
investigate what fraction of a cluster's stellar mass resides in the
BCG, the ICL, and in the satellite galaxies. We also study whether the
answer to this question depends on cluster mass. Finally, we examine
where the stars that end up in these three components at redshift
$z=0$ are formed, and in what objects they fall into the cluster.

\subsection{Intracluster light, satellite galaxies, and BCGs
in simulations and observations}
\label{sec:icl_fraction}

\subsubsection{The baryonic mass fractions}

\begin{figure}
\centerline{\includegraphics[width=\linewidth]
{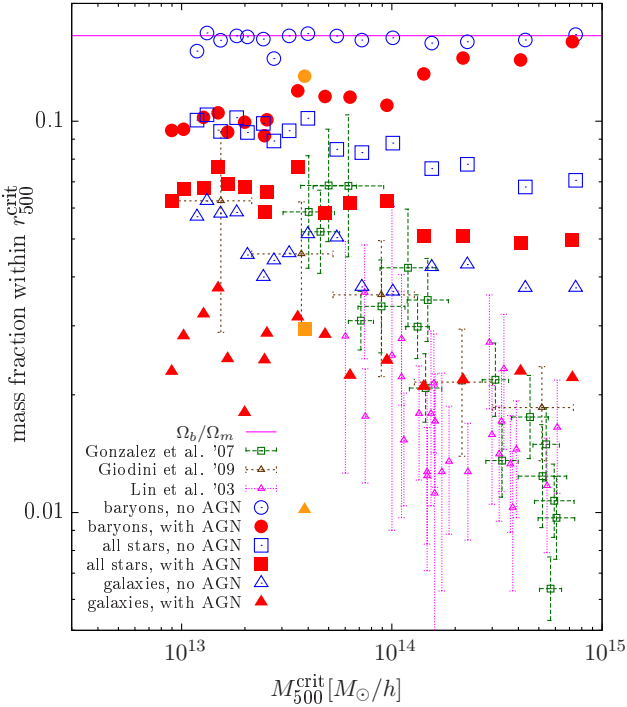}}
\caption{Mass fraction of baryons ({\it circles}), all stars
  (including ICL, {\it squares}), and stars residing in galaxies ({\it
    triangles}) within $r_{500}^\textrm{crit}$ for clusters and groups
  simulated with and without AGN feedback. The latter was found by
  summing up the stellar masses of the BCG, which was found by {\it
    Method 1}, and all satellite galaxies within
  $r_{500}^\textrm{crit}$. Observational data are shown for
  comparison. The \citet{Gonzalez2007} data should be compared to the
  mass fraction of all stars, while the other observational data
  should be compared to the mass fraction of stars found in
  galaxies. The cosmic baryon fraction assumed in the simulations is
  shown for reference. The {\it orange} symbols show results of a
  simulation where also kinetic supernova feedback was included.}
\label{fig:baryon_star_mass_frac}
\end{figure}

Figure~\ref{fig:baryon_star_mass_frac} illustrates the baryon and
stellar mass fractions of our simulated clusters, as well as the mass
fraction of stars that reside in cluster galaxies as a function of
cluster mass $M_{500}^\textrm{crit}$. Results from simulations both
with and without AGN feedback are shown, and for comparison with
observational constraints, data from \citet{Gonzalez2007},
\citet{Giodini2009} and \citet{Lin2003} are included.

In the simulations without AGN feedback, the baryon mass fractions
within $r_{500}^\textrm{crit}$ differ by less than $\sim10\%$ from the
assumed cosmic baryon fraction. When AGN are included, this holds only
for the most massive clusters, while there is a significant baryon
depletion in poor clusters and groups due to the AGN heating
\citep[see also][]{Puchwein2008}.

Without AGN, we obtain very large stellar mass fractions. In massive
clusters, roughly $40\%$ of the baryonic mass is found in stars, while
on the group scale this number rises to about $60\%$. This seems to be
inconsistent with observations, especially for massive clusters.  When
AGN are included, the stellar mass fractions within
$r_{500}^\textrm{crit}$ are lowered by about one third. One can
compare these total stellar mass fractions to the constraints from
\citet{Gonzalez2007}, as both cluster galaxies and intracluster stars
were accounted for in that study. The simulated and observed stellar
mass fractions are in good agreement on the group scale. However, we
do not find a comparably strong trend with halo mass in our
simulations as inferred by \citet{Gonzalez2007}, and accordingly we
obtain significantly larger stellar mass fractions for massive
clusters.

Also shown in Fig.~\ref{fig:baryon_star_mass_frac} is the mass
fraction of stars in cluster galaxies. More precisely, these values
were calculated by summing up the stellar masses of the BCG, which was
found by {\it Method 1}, i.e. using the radial cut-off scaled with the
cluster mass, and of all satellite galaxies within
$r_{500}^\textrm{crit}$. The resulting values can be compared to the
mass fractions of stars in cluster galaxies obtained by
\citet{Giodini2009} and \citet{Lin2003}. Without AGN, we again find a
significant discrepancy for massive clusters. Compared to the
observations the combined stellar mass of the cluster galaxies is too
large. Also on the group scale the mass fractions in the simulations
are larger than those from \citet{Lin2003}. However, they are roughly
consistent with the \citet{Giodini2009} results there. When AGN
feedback is included the combined stellar mass of the cluster galaxies
drops by about $\sim40\%$. This substantially improves the agreement
with observations for massive clusters. On the group scale, there is
then good agreement with the \citet{Lin2003} data, while
\citet{Giodini2009} inferred slightly larger mass fractions there.

Overall, the stellar mass fractions obtained without AGN feedback are
clearly too large. When the effects of AGN heating are included, the
stellar mass in cluster galaxies becomes however almost consistent
with observations. In particular, the total stellar mass,
i.e.~including the ICL, agrees with the \citet{Gonzalez2007} results
on the group scale. However, for massive clusters, we find a much
larger ICL contribution in the simulations than they infer from
observations. The simulations also do not fully reproduce the observed
strong trend of the stellar mass fractions with halo mass.

One should mention that we did not include kinetic supernova feedback
in the present set of simulations, instead all feedback from star
formation was injected thermally. Using supernova feedback that
includes a kinetic component can significantly reduce the amount
of stars formed in hydrodynamical cosmological simulations
\citep[e.g][]{Springel2003b,Borgani2006}. While this might help to
reproduce the overall stellar mass fractions in massive clusters, it
is, considering that such simulations \citep[see][]{Borgani2006} also
find only a weak trend of stellar mass fractions with halo mass,
unlikely to significantly improve the overall agreement with
observations.\footnote{Recently, more sophisticated kinetic supernova
  feedback schemes that tune the wind velocity and mass loading factor
  to properties of the host galaxy have been suggested
  \citep{Oppenheimer2006,Okamoto2009}. Such models might be more
  successful in reproducing a strong trend of the stellar mass
  fraction with halo mass.} Nevertheless, we
repeated the simulation of one of our clusters, this time including a
strong kinetic supernova feedback component,\footnote{We assumed a
supernova wind velocity of $\sim480$ km/s and a wind mass flux rate
that equals twice the star formation rate \citep[for details of the
wind model see][]{Springel2003}} in order to assess how it changes
the fraction of stars found in the BCG, the ICL, and in the satellite
galaxies. This run also included AGN feedback. Its results are
indicated by {\it orange} symbols in
Fig.~\ref{fig:baryon_star_mass_frac}. The corresponding simulation
without winds, is the one indicated by the red symbols at almost the
same (slightly lower) $M_{500}^\textrm{crit}$.

\subsubsection{The stellar fractions in the BCG, ICL, and satellite
  galaxies}
\label{sec:bcg_icl_sat_frac}

\begin{figure}
\centerline{\includegraphics[width=\linewidth]
{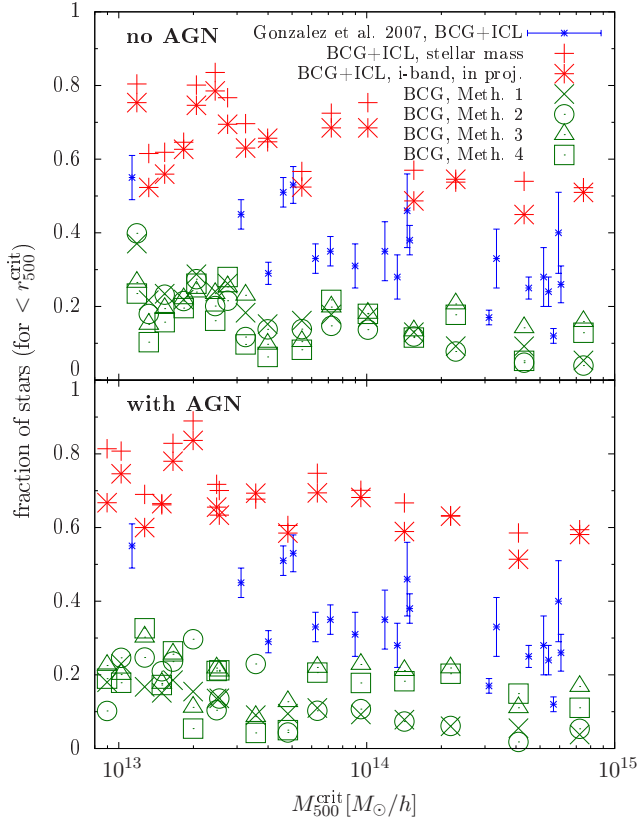}}
\caption{Fraction of the stellar mass within
$r_{500}^\textrm{crit}$ found in the BCG, and in the BCG+ICL component.
Mass fractions are shown as a function of cluster mass
$M_{500}^\textrm{crit}$ and for the four different methods of
determining a BCG mass (see Sect.~\ref{sec:identify_icl}). The {\it
upper panel} shows the values computed from runs without AGN, while the
{\it lower panel} illustrates the results obtained when including the
AGN feedback. Also shown is the BCG+ICL i-band luminosity fraction
within a projected radius on the sky (rather than 3D radius) of
$r_{500}^\textrm{crit}$. This data can be directly compared to the
observed BCG+ICL luminosity fractions from \citet{Gonzalez2007}.}
\label{fig:component_fractions}
\end{figure}

In the following, we discuss what fraction of stars in our simulated
clusters and groups reside in the BCG, the ICL, and in the satellite
galaxies, and how this compares to
observations. Fig.~\ref{fig:component_fractions} shows the mass
fractions of the stars found in the BCG and in the whole main halo,
i.e.~in the BCG+ICL component. The values were calculated within
$r_{500}^\textrm{crit}$. Results obtained from simulations both with
and without AGN are shown. The BCG fractions are reported for all the
four different methods we used to determine a BCG mass (see
\index{}Sect.~\ref{sec:identify_icl}). For reference, observational
constraints on the BCG+ICL luminosity fraction from
\citet{Gonzalez2007} are also shown. They should be compared to the
simulated BCG+ICL luminosity fractions indicated in the plot. The
latter were computed within a projected radius (rather than a 3D
radius) of $r_{500}^\textrm{crit}$, using the same method to obtain
stellar luminosities as described in Sect.~\ref{sec:identify_icl}.

Without AGN, we obtain BCG+ICL fractions of roughly $\sim50\%$ for
massive clusters, and of about $\sim70\%$ for groups. Note, however,
that these values are sensitive to the radius inside which they are
measured, because the distribution of the BCG+ICL component is more
concentrated than the distribution of the satellite galaxies. For
example, within $r_{200}^\textrm{crit}$ the values drop to about
$\sim45\%$ and $\sim65\%$, respectively. Accordingly, the BCG+ICL
fractions are also lower when calculated within a projected radius
rather than a 3D radius equal to $r_{500}^\textrm{crit}$. However, the
BCG+ICL luminosity fractions indicated in the figure are also affected
by small differences in the mass-to-light ratio. Namely, we find
mass-to-light ratios for the BCG+ICL components that are a few percent
smaller (up to ~10\% smaller for massive cluster, while being almost
identical for groups) than those of the whole stellar population in
the simulated halo. Thus, the i-band luminosity fractions are only
slightly smaller than the mass fractions calculated within a 3D radius
equal to $r_{500}^\textrm{crit}$. When accounting for these effects,
we nevertheless still obtain larger BCG+ICL fractions than derived
from the observations of \citet{Gonzalez2007}, especially for massive
clusters.

When AGN feedback is included, the stellar mass of the BCG+ICL
component decreases by about $\sim25\%$. However, the total stellar
masses decrease somewhat more strongly (see
Fig.~\ref{fig:baryon_star_mass_frac}). Thus, we find even slightly
larger BCG+ICL fractions in the simulations with AGN. A potential
explanation for this effect is that galaxies falling into the cluster
in simulations with AGN may be more prone to tidal disruption due to
their lower stellar mass and due to an expansion of their halos as a
consequence of the removal of gas by AGN heating.

Especially in the runs without AGN, there are several clusters for
which all the four different analysis methods for identifying the BCG
yield very similar BCG fractions and masses (see also
Fig.~\ref{fig:sb_profiles}).  There are, on the other hand, also
clusters where the results differ. In most of them, however, there is
still good agreement between the two methods that were inspired by
observational approaches, i.e.~between {\it Method 1}, where a radial
cut-off is used, and {\it Method 2}, where surface brightness profiles
are analyzed. Also, the methods that rely on the dynamics, i.e.~{\it
  Method 3} and {\it Method 4}, usually agree well with each other,
even though this is perhaps not too surprising as they both are based
on the same fit of the stellar velocity distribution. In the rare
cases were they do disagree, it is typically in unrelaxed objects,
e.g.~when an infalling group perturbs the stellar velocity
distribution and prevents an accurate two-component Maxwell fit.

Overall, there is hence some uncertainty in the BCG mass due to the
choice of the analysis method used for separating the BCG from the
ICL. However, for all our different methods only a small part of the
stars in the main halo, i.e.~in the BCG+ICL component, may be assigned
to the BCG.  This also means that independent of how exactly we choose
to make a distinction between BCG and ICL, we find a very significant
ICL component in our simulated clusters and groups.  We also note that
the BCG and ICL components in \cite{Gonzalez2005} are defined very
similarly to our {\it Method 2}, hence our analysis method should
yield results that can be directly compared to observations.

The fraction of stars in the ICL can be read off form the difference
between the BCG+ICL and BCG fractions in
Fig.~\ref{fig:component_fractions}. As can be seen, the ICL fractions
in our simulations are almost independent of the cluster mass. Within
$r_{500}^\textrm{crit}$, they are roughly $\sim45\%$ without AGN and
$\sim50\%$ with AGN physics. If we go to the larger radius of
$r_{200}^\textrm{crit}$ instead, they are slightly lower, that is
about $\sim40\%$ without AGN and $\sim45\%$ with AGN.

At a first glance these findings seem to be at odds with the results
of \cite{Murante2007}, where an ICL fraction rising with cluster mass
was found, from $\sim15\%$ at $10^{14} \, h^{-1} M_\odot$ to $\sim30\%$
at
$10^{15} \, h^{-1} M_\odot$. However, the lower values they obtained are
most
likely a consequence of the much lower numerical resolution in their
simulations, as they explicitly exclude stars from the ICL that got
liberated from poorly resolved galaxies. Looking only at the three
clusters they have simulated at high resolution, i.e.~with a mass
resolution comparable to our simulations, there is no significant
discrepancy. For those three objects they find ICL fractions of
$37\%$, $28\%$, and $41\%$, and no strong indication of a dependence
on cluster mass.

\subsubsection{The radial distribution of BCG+ICL stars}

\begin{figure}
\centerline{\includegraphics[width=\linewidth] {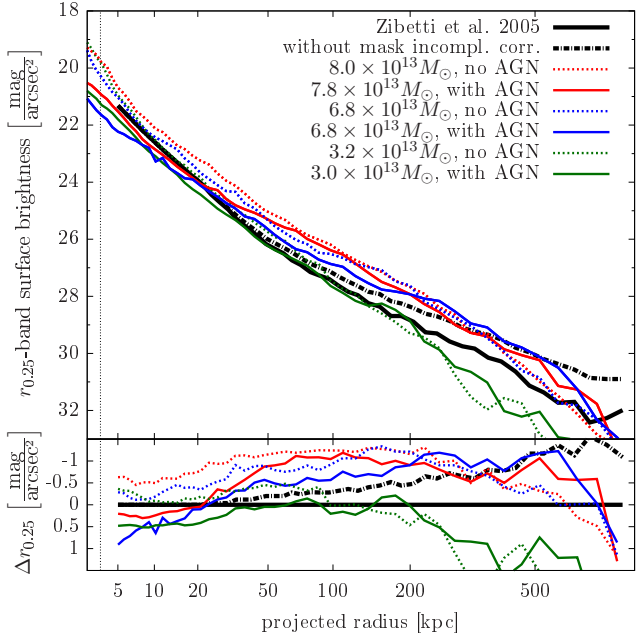}}
\caption{BCG+ICL surface brightness profiles of three
groups simulated with and without AGN feedback are shown in the {\it
upper panel}. For comparison
observational constraints from \citet{Zibetti2005} obtained by
stacking SDSS clusters and groups are shown. Also shown is the mean
surface brightness profile they find without correcting for incomplete
masking of satellite galaxies (assuming a constant fraction of
$\sim15\%$ of unmasked galaxy light). For each simulated halo, the mass
$M_{200}^\textrm{mean}$ is given in the figure's legend. The two
groups shown in {\it red} and {\it blue} have masses very similar to
the mean halo mass in \citet{Zibetti2005}. The {\it lower panel}
shows the difference to the profile obtained in \citet{Zibetti2005}. The
{\it vertical dotted line} indicates the gravitational softening.}
\label{fig:sb_comp}
\end{figure}

Fig.~\ref{fig:sb_comp} compares the surface brightness profiles of the
BCG+ICL component in three of our simulated groups to observational
constraints from \citet{Zibetti2005}.  For our simulated halos they
were calculated for simulation outputs at redshift $z=0.21$ of runs
with and without AGN, scaled to the reference redshift of
0.25.\footnote{The $r_{0.25}$-band of the photometric system defined
  in \citet{Zibetti2005} is used for this comparison.} The profile
from \citet{Zibetti2005} is a mean surface brightness profile obtained
by stacking SDSS cluster and group images, with a mean halo mass of
$M^\text{mean}_{200}\approx7 - 8\times10^{13} M_\odot$.  Also shown is
the profile they obtain without assuming a correction for incompletely
masked satellite galaxies.  In the figure, we compare with the two
simulated groups from our sample whose mass $M^\text{mean}_{200}$ is
closest to the mean mass of the observational sample.  The mass of
each simulated group is indicated in the figure's legend. Due to the
removal of gas by AGN heating, the mass in the runs with AGN is
typically slightly lower. In addition, we include results for a
further, lower mass group.

Without AGN, the surface brightness is typically too high in the
central regions.  In other words, the BCGs are overluminous in such
simulations. When AGN are included, the central profile of one of the
two groups, whose mass is close to the mean mass of the stacked
clusters and groups, agrees very well with the observed central
profile, indicating that a realistic BCG is formed in this
simulation. The other massive group shown in the plot has a somewhat
flatter central profile. However, given that there is object-to-object
scatter, no perfect agreement for all halos close to the mean mass is
expected.

At radii larger than about 30 kpc, the surface brightness of the two
more massive simulated groups exceeds the observed profile by roughly
$\sim 1 \, \textrm{mag}/\textrm{arcsec}^2$ or somewhat less when AGN
are included. This is consistent with our previous findings that we
have a very significant ICL component in our simulated halos, which as
this comparison confirms appears to exceed the observed amount of ICL.
As illustrated further in the figure, we would need to compare to a
simulated group of significantly lower mass to get a surface
brightness profile that is close to the mean observed one within 200
kpc.  However, in this case the simulated profile would then start to
significantly underpredict the surface brightness at even larger
radii. One should keep in mind though that for large radii this
comparison may be quite sensitive to the accuracy of the correction
for incomplete masking of satellite galaxies in \citet{Zibetti2005}.

\subsubsection{How do AGN affect the central cluster galaxies?}

Without a strong feedback mechanism, hydrodynamical simulations of
galaxy clusters suffer from excessive overcooling in cluster
cores. Typically, the central galaxy is fed by a strong cooling flow
and becomes too massive, too luminous and too blue compared to
observations. One of the main aims of including AGN heating in cluster
simulations was to offset cooling in cluster cores and produce more
realistic central galaxies. Comparing the BCGs in runs with and
without AGN feedback to observations should thus allow us to assess
whether the employed AGN feedback model successfully regulates the
central cooling flow.

\begin{figure}
\centerline{\includegraphics[width=\linewidth]
{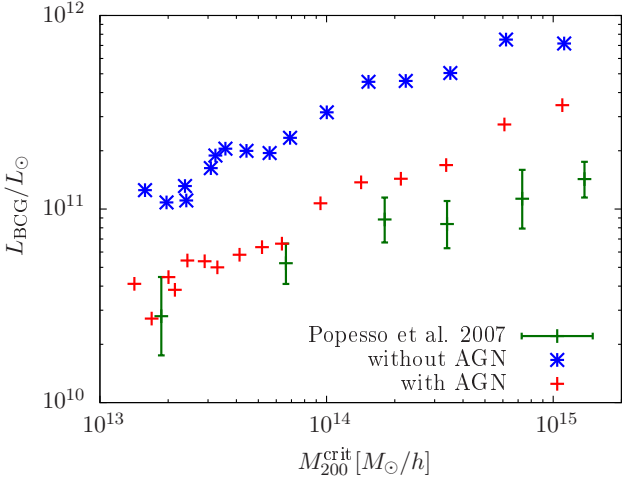}}
\caption{r-band BCG luminosity as a function of
cluster mass $M_{200}^\textrm{crit}$. Results are shown for runs with
and without AGN feedback. {\it Method 1} (see
Sect.~\ref{sec:identify_icl}) was used for finding the BCGs.
Observational data (median BCG luminosities for six cluster mass bins)
from \citet{Popesso2007} are shown for comparison. The AGN feedback
significantly reduces the BCG luminosities and improves agreement with
observations.}
\label{fig:bcg_lum_m200}
\end{figure}

Fig.~\ref{fig:bcg_lum_m200} shows the r-band luminosities of the BCGs
in our simulations as a function of cluster mass. Results are shown
for runs with and without AGN feedback and are compared to
observational constraints from \citet{Popesso2007}. Here {\it Method
  1} was used for finding the BCGs. The AGN heating successfully
reduces the BCG luminosities for all halo masses and strongly improves
the agreement with observations. For massive clusters the BCG
luminosities obtained with AGN are still somewhat larger than
observed. However, one should note that this depends somewhat on the
details of the method used for making a distinction between the BCG
and ICL components in the simulations. When {\it Method 3} or {\it 4}
is used, the resulting relation has much more scatter and the
agreement with observations in runs with AGN is not as good as with
{\it Method 1} or {\it 2}. Note that the reduced BCG mass in
our runs with AGN also reduces the gravitational lensing efficiency of
our galaxy clusters \citep[see][]{Mead2010}.

\subsection{When and where do the stellar populations of satellite
galaxies, BCGs, and the ICL form?}
\label{sec:formation_halos}

We now explore in what objects the stars have formed that are part of
satellite galaxies, the BCG, and the ICL at redshift $z=0$. Using the
IDs of the simulation particles, we can easily trace back every star
particle in the simulation to the first snapshot after its
formation. We then consider the {\small SUBFIND} (sub-)halo that the
star particle is part of in this snapshot as its formation
halo. Because we have saved simulation snapshots very frequently,
there is only a negligibly small number of particles that are not
bound to any halo in the first snapshot after their creation. We can
thus, in this way, assign formation halos to almost all star particles
in the simulation.

\begin{figure}
\centerline{\includegraphics[width=\linewidth]
{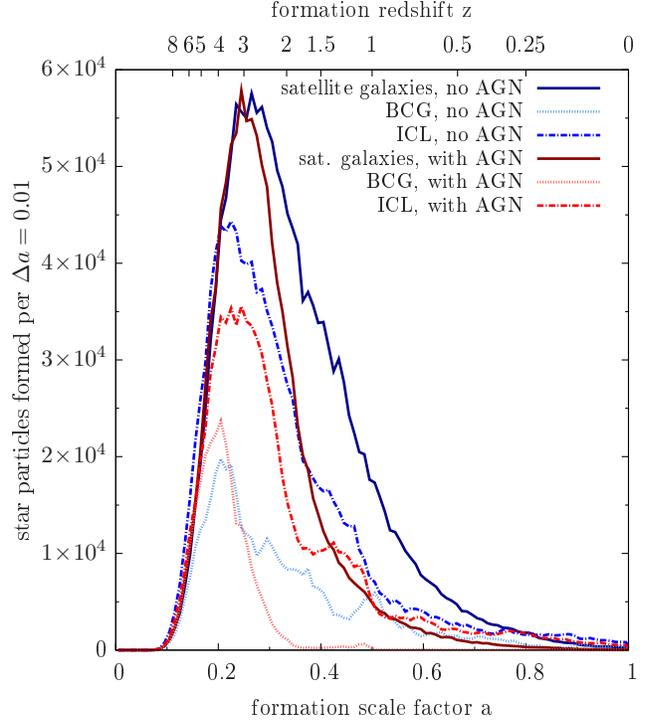}}
\caption{Distribution of the formation times of the stars in the
satellite galaxies, the BCG, and the ICL of a $10^{14} \, h^{-1}
M_\odot$
cluster. Results are shown for simulations with and without AGN
feedback. {\it Method 4} was used for making a distinction between BCG
and ICL. The simulations were performed at zoom factor 3 resolution (see
Table \ref{tab:resolutions}).}
\label{fig:form_times}
\end{figure}

Fig.~\ref{fig:form_times} shows the formation time of the star
particles that are found in the BCG, the ICL, and the satellite member
galaxies of a $10^{14} \, h^{-1} M_\odot$ cluster at $z=0$. Results are
shown
for simulations of this cluster with and without AGN feedback.
Comparing the two runs clearly shows that AGN efficiently suppress
star formation at low redshift. The effect is particularly strong for
the BCG where star formation is almost completely shut off at
redshifts $z<2$.  It is also interesting that in the run with AGN the
BCG stars form on average earlier than the stars in the satellite
galaxies, which nicely agrees with the observed cosmic
``down-sizing'', where the mass of galaxies hosting star formation
decreases with time \citep[e.g.][]{Cowie1996}. At first, this
observational finding is counterintuitive in a hierarchical structure
formation scenario.  However, using semi-analytic galaxy formation
models, \citet{DeLucia2006} have shown that AGN feedback allows
reproducing this behaviour in a $\Lambda$CDM cosmology where structure
forms hierarchically. Our results indicate that the AGN feedback model
in our cosmological hydrodynamical simulations operates in the same
way.

\begin{figure}
\centerline{\includegraphics[width=\linewidth]
{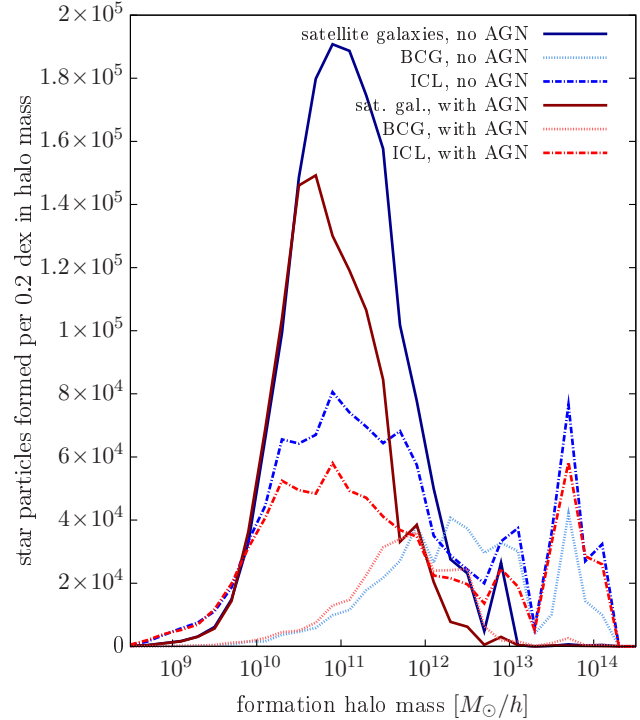}}
\caption{Distribution of the masses of the halos in which the stars in
the satellite galaxies, the BCG, and the ICL of a $10^{14} \,
h^{-1} M_\odot$
cluster formed. Results are shown for simulations with and without AGN
feedback. {\it Method 4} was used for making a distinction between BCG
and ICL. The simulations were performed at zoom factor 3 resolution (see
Table \ref{tab:resolutions}).}
\label{fig:form_masses}
\end{figure}

We also investigated in what objects stars form that at $z=0$ reside
in a cluster's satellite galaxies, in the BCG, or the
ICL. Fig.~\ref{fig:form_masses} shows the distribution of these stars
according to the mass of the halo in which they have formed. Here, the
halo mass is defined as the mass of the corresponding {\small SUBFIND}
(sub-)halo. For the central galaxy of a halo this is roughly the same
as $M^\text{crit}_{200}$, but for satellite galaxies it corresponds to
the smaller mass still bound within the gravitational tidal radius.
The measured distributions are shown for a $M^\text{crit}_{200} =
10^{14} \, h^{-1} M_\odot$ cluster and for runs with and without AGN.

Comparing the curves for the satellite galaxies, we can clearly see
that AGN suppress star formation in halos larger than $\sim
5\times10^{10} \, h^{-1} M_\odot$. Note that for smaller halos we do not
expect any difference, as in our simulations black holes are only
seeded in halos that exceed this threshold mass
\citep[see][]{Puchwein2008}. This also means that in the runs with AGN
the exact shape of the distribution at the low mass end will depend on
the adopted threshold for the seeding.

The stars that end up in the BCG at $z=0$ form in significantly more
massive halos. This is not only due to star formation in the massive
BCG itself, as for this cluster only $\sim 10\%$ and $\sim 30\%$ of
the stars that reside in the BCG at $z=0$ have formed in the BCG's
main progenitor in runs with and without AGN, respectively.  Instead,
also the stars that are acquired during mergers with the BCG tend to
be formed in more massive halos. This is not unexpected, however,
since the most massive galaxies are most likely to merge with the BCG
due to their shorter dynamical friction timescale. We shall return to
this point in Sect.~\ref{sec:fall_in_halos}. Without AGN, there is a
significant fraction of BCG stars that form in very massive halos,
i.e.  in halos with masses larger than $10^{13} \, h^{-1} M_\odot$.
Indeed,
many of these stars form in the cluster's main progenitor. On the
other hand, in the run with AGN, star formation is almost completely
shut off in the central galaxies of such massive halos, which again
demonstrates that our model for AGN feedback can efficiently suppress
strong cooling flows and excessive star formation in cluster cores.

The curves for the ICL peak at roughly the same halo mass as for the
satellite galaxies. However, the halo mass distribution is somewhat
broader. There is also a second peak at very high halo mass, which is
mostly due to stars formed in the cluster's main progenitor.  We find
a similar peak for most of our simulated clusters, with a position
that depends on the cluster's mass. The occurrence of this second peak
is rather surprising, especially since it is also found in runs with
AGN, in which there is basically no star formation in the BCG once the
cluster is as massive as required by the position of the peak. This
means that these stars are not formed in the BCG but somewhere else in
the cluster's main halo.

\begin{figure}
\centerline{\includegraphics[width=\linewidth]
{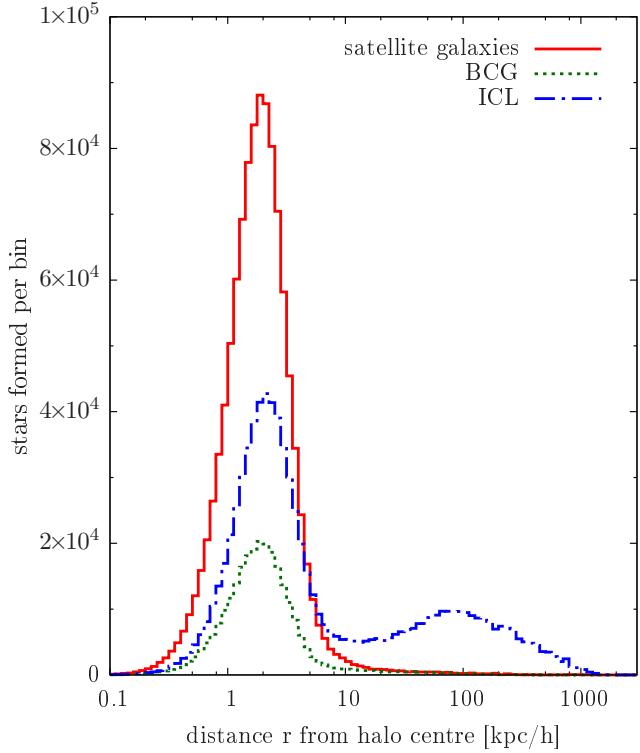}}
\caption{Distance of newly formed stars from the centres of the halos in
which they were formed. Distributions are given for stars that at $z=0$
end up in satellite galaxies, the BCG, and the ICL of a $10^{14}
\, h^{-1}
M_\odot$ cluster. {\it Method 4} was used for making a distinction
between BCG and ICL. Interestingly, a significant fraction of
intracluster stars is not formed in halo centres. The increase
of all distributions with distance for $r$ values smaller than $\sim 1\,
h^{-1}\textrm{kpc}$ is due to the increasing volume per radial bin.}
\label{fig:form_radius}
\end{figure}

This is illustrated for the same cluster in more detail in
Fig.~\ref{fig:form_radius}, which shows the distance of newly formed
stars from the centres of their host halos. Stars that at $z=0$ are
part of satellite galaxies or the BCG exclusively form in halo
centres, roughly within $10\,h^{-1}{\rm kpc}$. However, a
significant fraction ($\sim 30\%$) of stars that are part of the ICL at
$z=0$ form at much
larger distances between $10\,h^{-1}{\rm kpc}$ and $1\,h^{-1}{\rm
  Mpc}$. We found a similar fraction also in our other clusters.

We have investigated the nature of this `intracluster star formation'
in more detail, finding that it happens in small gas clouds that are
distributed throughout the cluster, which can in fact be easily
spotted in gas surface density maps. For most of these clouds,
however, an associated halo can not be identified in maps of the dark
matter particle distribution. We also checked whether these clouds
form by thermal instability from the intracluster medium or whether
they are stripped from infalling halos. We did this by tracing both
the star-forming gas particles in these clouds, as well as all gas
particles in the clusters main halo back to earlier times.  This
revealed that contrary to other gas particles, the vast majority of
star-forming gas particles fell into the cluster very recently. For
example, less than $5\%$ of the star-forming gas particles in these
clouds in the progenitor of a $10^{14} \, h^{-1} M_\odot$ cluster at
$z=1$ are
already part of this progenitor's main halo at $z=1.4$, while this
fraction is over $50\%$ when looking at all main halo gas particles at
the same cluster-centric radius. Also, at this earlier time most of
the gas particles that later form intracluster stars can be
unambiguously associated with infalling halos.

In other words, this means that most of these star-forming gas clouds
consist of material that was stripped out of small infalling
halos. Curiously, some of the gas clouds survived this stripping
basically intact and continue to form stars, which then become part of
the ICL.  Note that there is observational evidence that some star
formation does happen in gas stripped from infalling galaxies
\citep[e.g.][]{Sun2010}. Nevertheless, it is not entirely clear how
realistic this mode of `intracluster star formation' in our
simulations is, and whether it is in part occurring due to numerical
effects.  For example, SPH simulations are known to poorly resolve
fluid instabilities \citep[e.g.][]{Agertz2007} that could act to
disrupt such gas clouds after they are stripped from an infalling
halo.  Also, thermal conduction in the hot intracluster medium is not
included in our simulations and might play some role in reality,
unless it is very efficiently suppressed by magnetic fields.

If our simulations strongly overestimate the amount of such
`intracluster star formation', part of the discrepancy between our
simulation results and the amount of ICL inferred from observational
studies like \cite{Zibetti2005} and \cite{Gonzalez2007} could be
explained. In the most extreme case all star formation outside halo
centres may be spurious, in which case we would expect about
$\sim30\%$ less ICL than reported above. On the other hand, it is also
worth noting that if real clusters contain as many intracluster stars
as predicted by our simulations, this would significantly relax the
tension between constraints on the cosmic baryon fraction and baryon
fractions measured in galaxy clusters as, e.g., reported in
\cite{McCarthy2007}. Given the difficult observational challenge to
determine the amount of ICL accurately, it certainly appears possible
that the present observations have still missed a significant amount
of the ICL.

\subsection{In what objects do stars fall into the
forming cluster?}
\label{sec:fall_in_halos}

For each star particle that was not formed in the cluster's main
progenitor we try to find the galaxy in which it fell into the forming
cluster. We do this by looking for the most massive subhalo that a
star particle was part of before becoming part of the main cluster's
FoF group. This allows a determination of the properties of these
halos before tidal stripping in the cluster potential strongly affects
them.

\begin{figure}
\centerline{\includegraphics[width=\linewidth]
{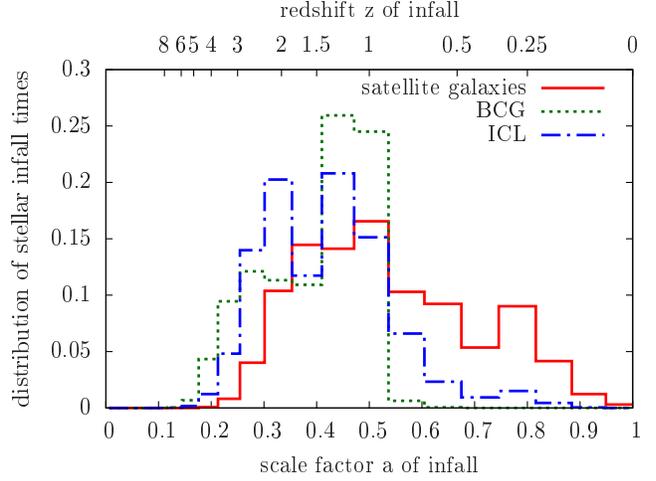}}
\caption{Distribution (fraction per bin, each histogram is normalized
  to unity) of stars according to the time of their infall into the
  forming cluster. Results are shown for stars assigned to satellite
  galaxies, the BCG, and the ICL at redshift $z=0$. {\it Method 4} was
  used for making a distinction between BCG and ICL. All curves were
  averaged over four clusters that were simulated with AGN. The same
  weight was used for each of them.}
\label{fig:infall_times}
\end{figure}

Fig.~\ref{fig:infall_times} shows the distribution of the stars in
satellite galaxy, BCG, and ICL components that were not formed in the
cluster's main progenitor according to the time of their infall into
the cluster. To reduce noise in the distributions we averaged them
over four clusters, assigning the same weight to each of them. The
simulations that were used for this figure included AGN feedback, but
we note that simulations without AGN yield very similar results. The
most obvious difference between the distributions is that almost all
stars that end up in the BCG or ICL fall into the cluster before
redshift $z\sim1$, while the vast majority of stars that fall into the
cluster later remain bound to their host galaxies. This suggests
that galaxies falling into the cluster after redshift $z\sim1$ do
usually not have enough time to sink towards the cluster
centre by dynamical friction and are thus unlikely to merge with the
BCG or get tidally disrupted close to the centre. This further
suggests that it takes a significant amount of time after the infall of
a galaxy into a cluster until stars are efficiently stripped from
it. Keeping this in mind, it is not at all surprising that
\cite{Murante2007} find that most of the stripping happens at $z<1$.

\begin{figure}
\centerline{\includegraphics[width=\linewidth]
{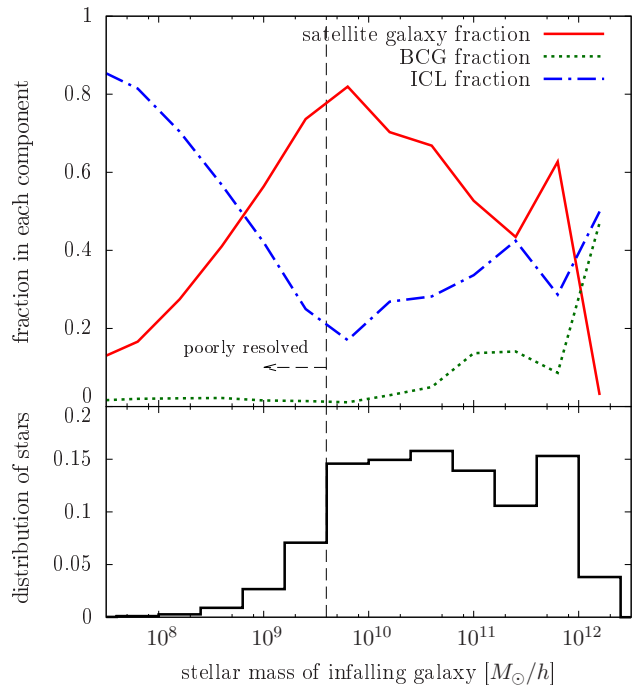}}
\caption{Distribution (fraction per bin) of stars according to the
  stellar mass of the galaxy in which they fell into the forming
  cluster ({\it lower panel}). The {\it upper panel} shows for each
  stellar mass bin of the infalling galaxies the fraction of stars
  that end up in satellite galaxies, the BCG, and the ICL at redshift
  $z=0$. {\it Method 4} was used for making a distinction between BCG
  and ICL. All curves were averaged over four clusters that were
  simulated with AGN. The same weight was used for each of them.}
\label{fig:infall_fracs}
\end{figure}

We also checked the stellar masses of the galaxies in which the
different components (satellite galaxy, BCG, and ICL stars) resided
when they fell into the forming cluster. This is illustrated in
Fig.~\ref{fig:infall_fracs}. The lower panel shows the distribution of
all infalling stars according to the stellar mass of the galaxy in
which they fell into the cluster. The upper panel shows for each
stellar mass bin of the infalling galaxy the fraction of stars that at
$z=0$ reside in the cluster's satellite galaxies, the BCG, and the
ICL. Again, the results were averaged over four clusters to reduce
noise in the curves. For stellar masses below $\sim 4\times 10^9
\, h^{-1} M_\odot$, the fraction of stars that get stripped from an
infalling
galaxy and become intracluster stars strongly increases in our
simulations. This is because these objects are poorly resolved, as
also shown by our analysis of the convergence of the properties of the
satellite galaxy population in Sect.~\ref{sec:res_studies}. However,
looking at the lower panel we see that only very few stars fall into
the cluster in such small, poorly resolved galaxies. Their disruption
does therefore not significantly bias our predictions for the ICL. On
the other hand, for more massive and well resolved galaxies, we find
that the fraction of stars that gets liberated and joins the ICL
increases with the galaxy's stellar mass. The most massive galaxies
are also most likely to merge with the BCG, as can be seen by the
strong increase of the BCG fraction. Accordingly, the fraction of
stars that remains bound to an infalling galaxy decreases with stellar
mass. This can be understood due to the shorter dynamical friction
timescale for massive galaxies, while at the same time their tidal
radius is larger.

Furthermore, we investigated what fraction of intracluster stars comes
from completely dissolved galaxies. We did this by constructing a
merger tree. For each {\small SUBFIND} (sub-)halo found in one
snapshot we searched for the descendant (sub-)halo in the next
snapshot that contains the largest number of the object's original
star particles. In case this would yield the cluster's main halo as
descendant, we also check which subhalo of the cluster contains the
largest number of the satellite galaxy's stars. If this subhalo still
contains more than $10\%$ of the original galaxy's star particles we
consider it as the descendant, and not the cluster's main halo. This
allows us to follow satellite galaxies even when they are strongly
tidally stripped between two subsequent simulation snapshots.

We find that the fraction of intracluster stars that comes from
dissolved galaxies, i.e.~galaxies that either merged with the BCG or
were completely tidally disrupted, depends on halo mass. While we have
performed this analysis only for four clusters, we find a systematic
trend where this fraction drops from $75\%$ for a $5\times 10^{13} \,
h^{-1} M_\odot$ group to $22\%$ for a $3\times 10^{14} \, h^{-1}
M_\odot$ cluster.

As discussed above, the mean fraction of stars that gets stripped from
an infalling galaxy and becomes part of the ICL depends on the mass of
the galaxy. On the other hand, this also means that the fraction of
intracluster stars may be biased in simulations that do not reproduce
the correct galaxy stellar mass function. Getting the latter right in
cosmological hydrodynamical simulations is, however, a long-standing
problem which has not yet been solved satisfactorily. Looking at the
upper panel of Fig.~\ref{fig:infall_fracs}, we see that we might
overestimate the ICL fraction if we significantly overpredict the
number of very low mass or very massive galaxies. The former does not
seem to be a problem as indicated by the distribution in the figure's
lower panel.  The abundance of massive galaxies, however, is typically
overpredicted in cosmological hydrodynamical simulations (see e.g.
\cite{Oppenheimer2009}). This surely is the case in our simulations
without AGN, but we can not be fully sure yet whether this is
completely resolved when the AGN feedback model is included.
Assessing this reliably requires a high resolution simulation of a
large cosmological box, which has not yet been done due to the large
computational cost.  

Nevertheless, from exploring the cluster galaxy mass functions of our
simulated clusters, we can be confident that the problem of
overpredicting massive galaxies is rather strongly alleviated in runs
with AGN, as also suggested by the much lower BCG luminosities shown
in Fig.~\ref{fig:bcg_lum_m200}.  However, the ICL fractions in our
simulations with AGN feedback are even slightly higher than in the
runs without it. This suggests that overpredicting massive galaxies is
not the main reason why our ICL fractions are relatively high compared
to the values typically inferred from observations. This also
highlights that it is really far from obvious how a large ICL fraction
can be avoided in the simulations of the formation of clusters in the
$\Lambda$CDM cosmology.

\subsection{The fate of infalling galaxies}
\label{sec:infalling_galaxies}

\begin{figure*}
\centerline{\includegraphics[width=\linewidth]
{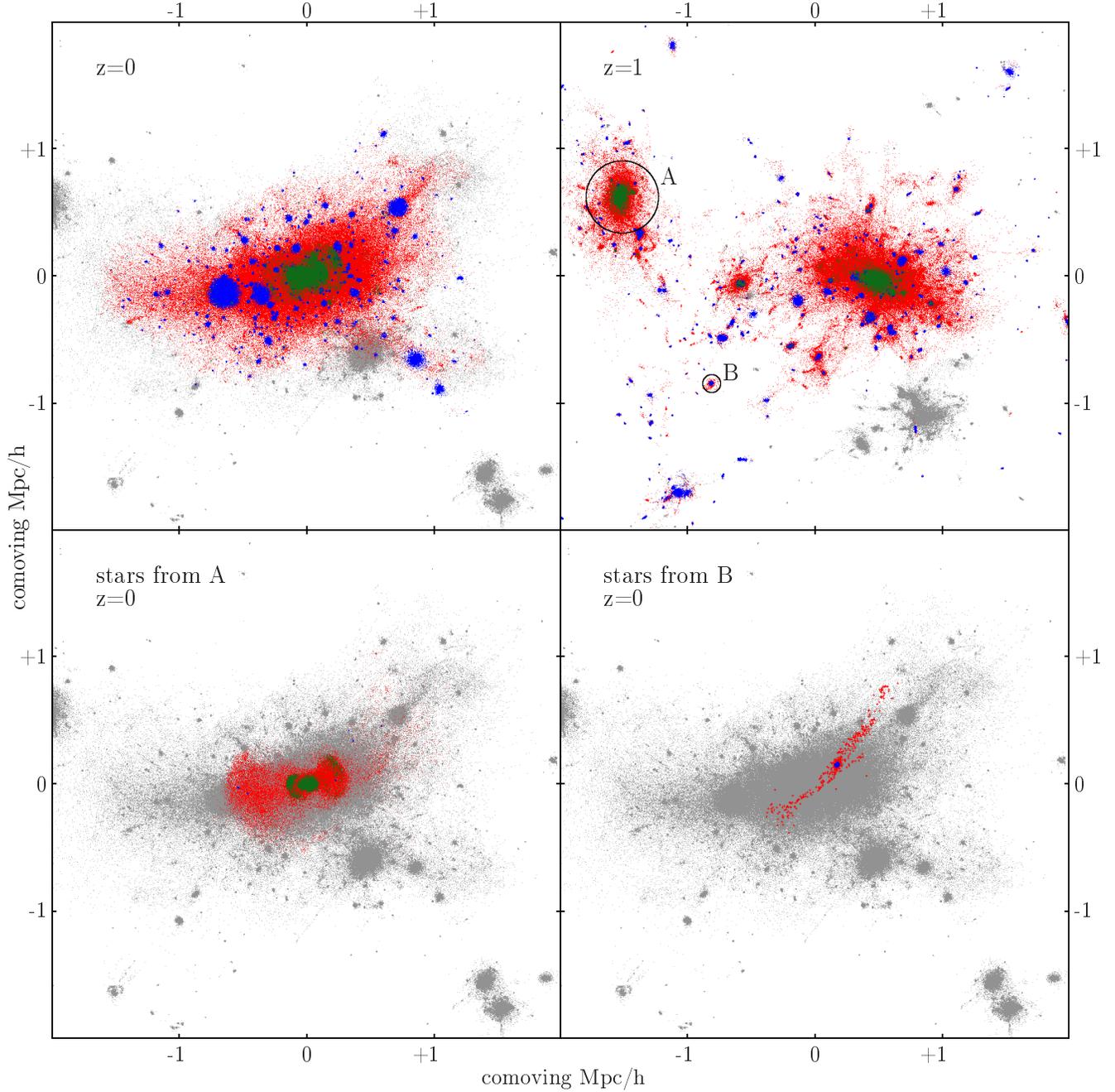}}
\caption{Distribution of stars in a $10^{14} \, h^{-1} M_\odot$ cluster
  simulated with AGN feedback. The {\it upper left panel} shows the
  stars residing in the cluster's satellite galaxies ({\it blue}), BCG
  ({\it green}), and ICL ({\it red}) at redshift $z=0$. {\it Grey} dots
  indicate stars that are not part of the clusters FoF group at
  $z=0$. {\it Method 4} was used for making a distinction between BCG
  and ICL. The {\it upper right panel} shows the cluster's progenitors
  at $z=1$. All star particles are coloured as before, i.e. according to
  the component in which they end up at $z=0$. The circles mark two
  infalling objects A and B. In the {\it lower panels} the positions
  of the stars that are part of A ({\it lower left}) and B ({\it lower
    right}) at $z=1$ are shown at $z=0$. Again they are coloured according
  to the component to which they are assigned at $z=0$. All stars that
  are not part of A or B are show in {\it grey} in the {\it lower
    panels}.}
\label{fig:clus_img}
\end{figure*}

Fig.~\ref{fig:clus_img} illustrates the infall of satellite galaxies
and the origin of intracluster stars for a $10^{14} \, h^{-1} M_\odot$
cluster. The upper left panel shows all star particles residing in the
cluster's satellite galaxies, and the BCG and ICL components at
$z=0$. Here, {\it Method 4} was used for making a distinction between
BCG and ICL, and the figure illustrates the assignment of the star
particles to the individual components. The upper right panel of
Fig.~\ref{fig:clus_img} shows the same cluster at $z=1$. All star
particle are still coloured according to the component which they
belong to at $z=0$, allowing some insights into the assembly history
of the cluster galaxy population and the origin of intracluster
stars. 

Looking at some of the infalling galaxies, it can be easily seen that
already at $z=1$ stars that later become part of the ICL are
preferentially found in the outskirts of infalling galaxies, and are
therefore comparatively weakly bound to them. Furthermore, the figure
shows that only the most massive infalling object (marked as A)
contributes a significant number of stars to the BCG, as expected from
our analysis in the previous section. Following the evolution of
clusters in this way also nicely shows that galaxy mergers often
create a loosely bound component of stars which is subsequently
stripped when the merger remnant falls into the
cluster. Unfortunately, it is a bit hard to illustrate this in a still
image, but it confirms the finding of \cite{Murante2007} that mergers
in the assembly history of the BCG and of other massive cluster
galaxies are of critical importance for the formation of the ICL.

We now consider the fate of some infalling galaxies in more
detail. For this purpose we have selected two infalling objects marked
as A and B in the upper right panel of Fig.~\ref{fig:clus_img} and
followed them to redshift $z=0$. The distribution of their stars at
$z=0$ is shown in the lower panels. The core of group-sized object A
has basically merged with the BCG at this point. However, even after
several passages of the cluster core one can still see significant
tidal features both in the distribution of stars that at $z=0$ are
assigned to the BCG and the ICL. The infalling galaxy B, on the
other hand, remains largely intact and becomes one of the cluster's
satellite galaxies at $z=0$.  Those stars that were stripped from it
form a large tidal stream extending almost over the whole
cluster. These two examples illustrate that the ICL really consists of
many individual tidal streams and shell-like features. Only their
superposition looks like a smooth distribution of intracluster stars.

\begin{figure}
\centerline{\includegraphics[width=\linewidth]
{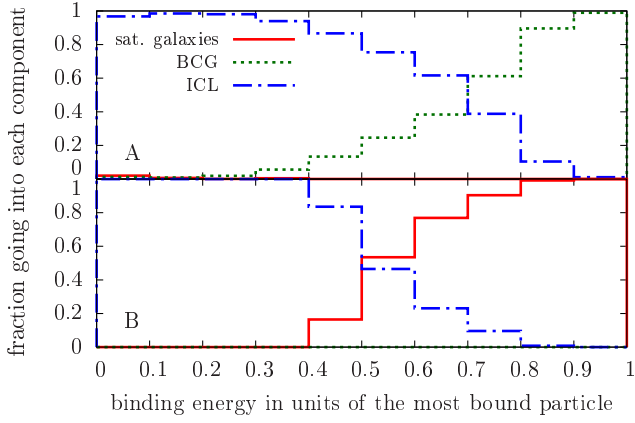}}
\caption{Distribution of binding energies in their host halos at $z=1$
  of the stars that end up the cluster's satellite galaxies, BCG, and
  ICL at $z=0$. Results are shown for the stars that are part of the
  objects A ({\it upper panel}) and B ({\it lower panel}) indicated in
  Fig.~\ref{fig:clus_img}. {\it Method 4} was used for making a
  distinction between BCG and ICL.}
\label{fig:binding_egy}
\end{figure}

As discussed above, stars that later become part of the ICL are
preferentially found in the outskirts of infalling galaxies (see also
the upper right panel of Fig.~\ref{fig:clus_img}). It also seems
plausible that the most weakly bound stars are the ones most likely to
be stripped from an infalling galaxy and to become intracluster
stars. We have checked this explicitly by calculating the binding
energy of stars in such infalling galaxies, and
Fig.~\ref{fig:binding_egy} illustrates the results of our analysis. We
show the fraction of stars that at $z=0$ reside in the cluster's
satellite galaxies, the BCG, and the ICL as a function of their
binding energy in objects A and B at $z=1$. As expected, the fraction
of stars that become part of the ICL strongly decreases with
increasing binding energy. For object A, the most bound stars end up
in the cluster's central galaxy, while for object B, the most bound
stars remain bound together in a satellite galaxy. Overall, about half
of the stars in A end up in the BCG, while the other half becomes part
of the ICL. For object B, only about $\sim 10\%$ of the stars are
stripped and become intracluster stars.

\begin{figure}
\centerline{\includegraphics[width=\linewidth]
{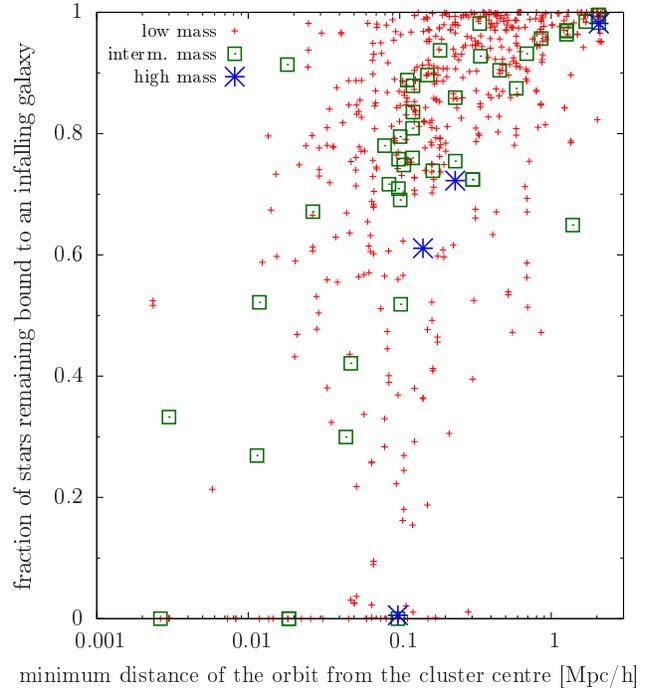}}
\caption{Fraction of stars that remain bound to a satellite galaxy as
  a function of the minimum distance of its orbit from the cluster
  centre. All galaxies that fell into the main progenitor of a
$3\times10^{14} \, h^{-1} M_\odot$
  cluster between redshift $z=1.5$ and $z=1$ are plotted. The minimum
  distance was found by interpolating the galaxy's orbit based on all
  simulation snapshots between the time of the infall and
  $z=0$. Different symbols are used for {\it low mass} galaxies (stellar
  mass before infall into the cluster $M_*$ in the range
  $5\times10^9 \, h^{-1} M_\odot \le M_* < 5\times10^{10}\,
h^{-1} M_\odot$), {\it
    intermediate mass} galaxies ($5\times10^{10} \, h^{-1} M_\odot \le
M_*
< 5\times10^{11} \, h^{-1} M_\odot$) and {\it high mass} galaxies ($M_*
\ge 5\times10^{11} \, h^{-1} M_\odot$). Galaxies with a fraction of 0
have either merged with the BCG or have been completely tidally
disrupted.}
\label{fig:bound_frac}
\end{figure}

We have shown in Fig.~\ref{fig:infall_fracs} that the fraction of
stars that gets stripped from an infalling galaxy depends on the
galaxy's mass. However, it also depends on the orbit of the
galaxy. This is illustrated in Fig.~\ref{fig:bound_frac}, which shows
the fraction of stars that at $z=0$ are still bound to a satellite
galaxy as a function of the minimum distance of the galaxy's orbit
from the cluster centre. The fractions are shown for all galaxies that
fell into the progenitor of a $3\times10^{14} \, h^{-1} M_\odot$ cluster
between redshifts $z=1.5$ and $z=1$. In order to determine the minimum
distance, the orbit of the galaxy and of the cluster centre were
interpolated using all simulation snapshots between the time of the
galaxy's infall and $z=0$, using a cubic interpolation between each
pair of successive snapshots based on the halo positions and
velocities determined by {\small SUBFIND}.  The results are shown for
three different stellar mass ranges of the infalling galaxy.  We see
that only few stars are stripped from galaxies that never come close
to the cluster centre. On the other hand, galaxies that closely
approach the cluster centre are often substantially stripped,
completely tidally disrupted, or merge with the BCG. This confirms the
expectation that most of the tidal interactions that liberate
intracluster stars happen close to the cluster centre and the BCG.

\section{Summary and conclusions}
\label{sec:summary}

The intracluster light in clusters of galaxies represents an interesting and
significant component of their total stellar mass. While the observational
constraints on the ICL are still uncertain, some studies have reported
intracluster star fractions of up to $\sim 50\%$. Even if the true fraction is
significantly lower, it is therefore clear that the ICL can not be neglected
in the baryon and stellar mass budgets of clusters. Furthermore, the radial
profile of the ICL and the mass contained in it may pose interesting
constraints on galaxy cluster formation models. Yet, comparatively little
theoretical work has been carried out thus far on the formation of the ICL. In
fact, most semi-analytic models of galaxy formation have hitherto
ignored this component entirely.

In this work we have therefore studied the origin of intracluster stars in a
set of high-resolution hydrodynamical simulations of the formation of clusters
of galaxies, embedded in their appropriate cosmological setting. Our sample of
resimulations has been randomly drawn from the Millennium simulation, with the
only selection criterion being to provide a wide coverage of group and cluster
masses, of roughly two orders of magnitude in halo mass. Thanks to the
very high
mass and force resolution of our resimulations, our simulations provide
a powerful and representative sample of the whole cluster population.
Another very timely aspect of our simulations is that we not only
account for hydrodynamics, radiative cooling, heating by a UV
background, star formation and supernovae feedback, but also
incorporate a state-of-the-art model for
the growth of supermassive black holes and for feedback from AGN.
Because we simulated each cluster both with and without AGN physics,
this allows us to pinpoint the impact of AGN heating on the cluster
galaxy populations, and in particular on the ICL.

Our results clearly confirm the importance of AGN feedback in galaxy clusters.
AGN lead to a reduction of the amount of stars in our clusters and groups by
about one third, roughly independent of cluster mass.  Especially the stellar
masses and luminosities of BCGs are greatly reduced by AGN feedback, and their
stellar populations become much older.  As a consequence, the BCGs are in much
better agreement with observational constraints.  Furthermore, in poor clusters
and groups, the total baryon fractions within $r^\text{crit}_{500}$
become significantly lower when AGN heating is included.

The primary focus of our analysis has been on the amount and the origin of the
ICL component in our simulated groups and clusters.  In order to allow a
meaningful comparison of our simulation results with observations of clusters
and groups, it was necessary to find a robust way to assign star particles in
the simulations either to one of the cluster's satellite galaxies, to the BCG,
or to the ICL.  Especially the distinction between the latter two components
is not without ambiguities. We therefore developed and tested several
different methods for distinguishing between the different components.  This
helps to assess the robustness of the measurements and allows the validation
of measurement techniques that are close to the methods usually employed in
observational studies.

Our investigation of the properties of the satellite galaxy populations, the
BCGs and the ICL components in our simulations yielded the following
main results:

\begin{itemize}

\item We find a very significant fraction of $\sim40\%$ (without AGN) or
  $\sim45\%$ (with AGN) of intracluster stars within $r^\text{crit}_{200}$ in
  our simulated clusters and groups.  These values are robust with respect to
  numerical resolution and integration accuracy, and are almost independent of
  halo mass.  They are, however, larger than those typically inferred from
  observations.

\item For all the different methods we tested for making a distinction between
  BCG and ICL, we find that the vast majority of stars in the main halo of our
  simulated clusters are part of the ICL, rather than the BCG.  As a result of
  this dominance, our ICL fractions also do not strongly depend on the exact
  method used for making the distinction.

\item The intracluster stars form on average earlier than the stars residing
  in cluster satellite galaxies. Stars ending up in the BCG at $z=0$, however,
  typically form at even earlier times, especially when AGN are included in
  the simulations, which prevent excessive star formation in BCGs at low
  redshift.

\item The fraction of stars that is stripped from infalling galaxies
  increases with galaxy stellar mass, which can be understood as a consequence
  of their larger ratio of size to tidal radius.  The orbit of an infalling
  galaxy also plays an important role.  Galaxies that closely approach the
  cluster centre typically loose the largest fraction of stars.  

\item We find that most intracluster stars are stripped from galaxies that
  fell into the cluster early, roughly before $z\sim1$. Most likely, galaxies
  falling in later just do not have enough time to sink towards the cluster
  centre by dynamical friction and to get tidally disrupted there.

\item The stars stripped from individual galaxies are not smoothly distributed
  within the cluster even after several orbits of their former host galaxy,
  instead they form streams and other tidal features. The ICL is therefore not
  smooth but a superposition of many such tidal features.

\item In our simulations, a significant fraction of intracluster stars (up to
  $\sim30\%$) do not form in galaxies but in cold gas clouds stripped from
  infalling substructures. These clouds remain intact after being stripped and
  give rise to `intracluster star formation'.

\item When including AGN, the total stellar masses in our simulated galaxy
  groups are in good agreement with observations. However, the simulations do
  not reproduce the steep observed decline of the stellar mass fraction with
  halo mass, so that our massive simulated clusters contain more stars than
  suggested by observations.

\end{itemize}

It will be very interesting to see whether improved observational
determinations of the ICL fraction in clusters confirm a lower value than
found in our simulations. This would represent a non-trivial challenge for
future simulation models, as our work has shown that the simulated ICL
fraction is remarkably robust, not only with respect to integration parameters
and numerical resolution, but also to quite drastic changes in the
modelling of star formation. This also means that it is not obvious
at all how a cluster simulation could be obtained that reproduces the
galaxy population as well as in our best runs but at the same time
yields a much lower ICL component.

One interesting effect we found in our simulations is that a sizable fraction
of our intracluster stars actually forms `in situ', in gas clouds that were
stripped out of the dark matter halos of infalling galaxies but that are not
bound to a dark matter satellite any more.  It is not entirely clear whether
this mode of `intracluster star formation' is significantly affected by
numerical inaccuracies in the simulations. For example, fluid instabilities
might be able to disrupt these clouds and suppress the star formation in them
if they are better resolved than possible with SPH.  If this were indeed the
case, this could explain some part of the discrepancy. On the other hand, a
high ICL fraction would actually help to relax the current tensions between
constraints on the cosmic baryon fraction and the baryon fractions measured in
galaxy clusters.

\section*{Acknowledgments}
We would like to thank Simon White for very constructive discussions.
D.S. acknowledges Postdoctoral Fellowship from the UK Science and
Technology Funding Council (STFC). K.D. acknowledges support by the
DFG Priority Programme 117. Part of the simulations have been
performed on the Cambridge High Performance Computing Cluster Darwin.

\appendix

\bibliographystyle{mnbst}
\bibliography{paper}

\begin{thebibliography}{38}
\providecommand{\natexlab}[1]{#1}

\bibitem[{{Agertz} et~al.(2007)}]{Agertz2007}
{Agertz} O., et~al., 2007, \mnras, 380, 963

\bibitem[{{Baldry} et~al.(2008){Baldry}, {Glazebrook}, and
  {Driver}}]{Baldry2008}
{Baldry} I.K., {Glazebrook} K., {Driver} S.P., 2008, \mnras, 388, 945

\bibitem[{{Bernardi} et~al.(2007){Bernardi}, {Hyde}, {Sheth}, {Miller}, and
  {Nichol}}]{Bernardi2007}
{Bernardi} M., {Hyde} J.B., {Sheth} R.K., {Miller} C.J., {Nichol} R.C., 2007,
  \aj, 133, 1741

\bibitem[{{Borgani} et~al.(2006)}]{Borgani2006}
{Borgani} S., et~al., 2006, \mnras, 367, 1641

\bibitem[{{Bruzual} and {Charlot}(2003)}]{Bruzual2003}
{Bruzual} G., {Charlot} S., 2003, \mnras, 344, 1000

\bibitem[{{Cowie} et~al.(1996){Cowie}, {Songaila}, {Hu}, and
  {Cohen}}]{Cowie1996}
{Cowie} L.L., {Songaila} A., {Hu} E.M., {Cohen} J.G., 1996, \aj, 112, 839

\bibitem[{{De Lucia} et~al.(2006){De Lucia}, {Springel}, {White}, {Croton}, and
  {Kauffmann}}]{DeLucia2006}
{De Lucia} G., {Springel} V., {White} S.D.M., {Croton} D., {Kauffmann} G.,
  2006, \mnras, 366, 499

\bibitem[{{de Vaucouleurs}(1948)}]{deVaucouleurs1948}
{de Vaucouleurs} G., 1948, Annales d'Astrophysique, 11, 247

\bibitem[{{Dolag} et~al.(2008){Dolag}, {Borgani}, {Murante}, and
  {Springel}}]{Dolag2008}
{Dolag} K., {Borgani} S., {Murante} G., {Springel} V., 2008, ArXiv e-prints:
  0808.3401

\bibitem[{{Dolag} et~al.(2009){Dolag}, {Murante}, and {Borgani}}]{Dolag2009}
{Dolag} K., {Murante} G., {Borgani} S., 2009, ArXiv e-prints: 0911.1129

\bibitem[{{Giodini} et~al.(2009)}]{Giodini2009}
{Giodini} S., et~al., 2009, \apj, 703, 982

\bibitem[{{Gonzalez} et~al.(2005){Gonzalez}, {Zabludoff}, and
  {Zaritsky}}]{Gonzalez2005}
{Gonzalez} A.H., {Zabludoff} A.I., {Zaritsky} D., 2005, \apj, 618, 195

\bibitem[{{Gonzalez} et~al.(2007){Gonzalez}, {Zaritsky}, and
  {Zabludoff}}]{Gonzalez2007}
{Gonzalez} A.H., {Zaritsky} D., {Zabludoff} A.I., 2007, \apj, 666, 147

\bibitem[{{Komatsu} et~al.(2008)}]{Komatsu2008}
{Komatsu} E., et~al., 2008, ArXiv e-prints: 0803.0547

\bibitem[{{Lin} and {Mohr}(2004)}]{Lin2004}
{Lin} Y., {Mohr} J.J., 2004, \apj, 617, 879

\bibitem[{{Lin} et~al.(2003){Lin}, {Mohr}, and {Stanford}}]{Lin2003}
{Lin} Y.T., {Mohr} J.J., {Stanford} S.A., 2003, \apj, 591, 749

\bibitem[{{McCarthy} et~al.(2007){McCarthy}, {Bower}, and
  {Balogh}}]{McCarthy2007}
{McCarthy} I.G., {Bower} R.G., {Balogh} M.L., 2007, \mnras, 377, 1457

\bibitem[{{McCarthy} et~al.(2009)}]{McCarthy2009}
{McCarthy} I.G., et~al., 2009, ArXiv e-prints: 0911.2641

\bibitem[{{Mead} et~al.(2010){Mead}, {King}, {Sijacki}, {Leonard}, {Puchwein},
  and {McCarthy}}]{Mead2010}
{Mead} J.M.G., {King} L.J., {Sijacki} D., {Leonard} A., {Puchwein} E.,
  {McCarthy} I.G., 2010, ArXiv e-prints: 1001.2281

\bibitem[{{Murante} et~al.(2007){Murante}, {Giovalli}, {Gerhard}, {Arnaboldi},
  {Borgani}, and {Dolag}}]{Murante2007}
{Murante} G., {Giovalli} M., {Gerhard} O., {Arnaboldi} M., {Borgani} S.,
  {Dolag} K., 2007, \mnras, 377, 2

\bibitem[{{Murante} et~al.(2004)}]{Murante2004}
{Murante} G., et~al., 2004, \apjl, 607, L83

\bibitem[{{Okamoto} et~al.(2009){Okamoto}, {Frenk}, {Jenkins}, and
  {Theuns}}]{Okamoto2009}
{Okamoto} T., {Frenk} C.S., {Jenkins} A., {Theuns} T., 2009, ArXiv e-prints:
  0909.0265

\bibitem[{{Oppenheimer} and {Dav{\'e}}(2006)}]{Oppenheimer2006}
{Oppenheimer} B.D., {Dav{\'e}} R., 2006, \mnras, 373, 1265

\bibitem[{{Oppenheimer} et~al.(2009){Oppenheimer}, {Dav{\'e}}, {Kere{\v s}},
  {Fardal}, {Katz}, {Kollmeier}, and {Weinberg}}]{Oppenheimer2009}
{Oppenheimer} B.D., {Dav{\'e}} R., {Kere{\v s}} D., {Fardal} M., {Katz} N.,
  {Kollmeier} J.A., {Weinberg} D.H., 2009, ArXiv e-prints: 0912.0519

\bibitem[{{Popesso} et~al.(2007){Popesso}, {Biviano}, {B{\"o}hringer}, and
  {Romaniello}}]{Popesso2007}
{Popesso} P., {Biviano} A., {B{\"o}hringer} H., {Romaniello} M., 2007, \aap,
  464, 451

\bibitem[{{Puchwein} et~al.(2008){Puchwein}, {Sijacki}, and
  {Springel}}]{Puchwein2008}
{Puchwein} E., {Sijacki} D., {Springel} V., 2008, \apjl, 687, L53

\bibitem[{{Schombert}(1986)}]{Schombert1986}
{Schombert} J.M., 1986, \apjs, 60, 603

\bibitem[{{Sijacki} et~al.(2007){Sijacki}, {Springel}, {di Matteo}, and
  {Hernquist}}]{Sijacki2007}
{Sijacki} D., {Springel} V., {di Matteo} T., {Hernquist} L., 2007, \mnras, 380,
  877

\bibitem[{{Sommer-Larsen} et~al.(2005){Sommer-Larsen}, {Romeo}, and
  {Portinari}}]{Sommer-Larsen2005}
{Sommer-Larsen} J., {Romeo} A.D., {Portinari} L., 2005, \mnras, 357, 478

\bibitem[{{Springel}(2005)}]{Springel2005c}
{Springel} V., 2005, \mnras, 364, 1105

\bibitem[{{Springel} and {Hernquist}(2003{\natexlab{a}})}]{Springel2003}
{Springel} V., {Hernquist} L., 2003{\natexlab{a}}, \mnras, 339, 289

\bibitem[{{Springel} and {Hernquist}(2003{\natexlab{b}})}]{Springel2003b}
{Springel} V., {Hernquist} L., 2003{\natexlab{b}}, \mnras, 339, 312

\bibitem[{{Springel} et~al.(2001){Springel}, {White}, {Tormen}, and
  {Kauffmann}}]{Springel2001}
{Springel} V., {White} S.D.M., {Tormen} G., {Kauffmann} G., 2001, \mnras, 328,
  726

\bibitem[{{Springel} et~al.(2005{\natexlab{a}}){Springel}, {Di Matteo}, and
  {Hernquist}}]{Springel2005b}
{Springel} V., {Di Matteo} T., {Hernquist} L., 2005{\natexlab{a}}, \mnras, 361,
  776

\bibitem[{{Springel} et~al.(2005{\natexlab{b}})}]{Springel2005a}
{Springel} V., et~al., 2005{\natexlab{b}}, \nat, 435, 629

\bibitem[{{Sun} et~al.(2010){Sun}, {Donahue}, {Roediger}, {Nulsen}, {Voit},
  {Sarazin}, {Forman}, and {Jones}}]{Sun2010}
{Sun} M., {Donahue} M., {Roediger} E., {Nulsen} P.E.J., {Voit} G.M., {Sarazin}
  C., {Forman} W., {Jones} C., 2010, \apj, 708, 946

\bibitem[{{Willman} et~al.(2004){Willman}, {Governato}, {Wadsley}, and
  {Quinn}}]{Willman2004}
{Willman} B., {Governato} F., {Wadsley} J., {Quinn} T., 2004, \mnras, 355, 159

\bibitem[{{Zibetti} et~al.(2005){Zibetti}, {White}, {Schneider}, and
  {Brinkmann}}]{Zibetti2005}
{Zibetti} S., {White} S.D.M., {Schneider} D.P., {Brinkmann} J., 2005, \mnras,
  358, 949

\end{thebibliography}
\end{document}